\documentclass[twocolumn]{aastex63}

\usepackage{gensymb,xcolor}
\usepackage{lineno}
\defcitealias{riechers2011}{R11}
\received{\today}
\submitjournal{ApJ}

\shorttitle{SMM J13120+4242}
\shortauthors{Frias Castillo et al.}

\graphicspath{{./}{figures/}}

\begin{document}

\title{Kiloparsec-scale imaging of the CO(1-0)-traced cold molecular gas reservoir in a z$\sim$3.4 submillimeter galaxy}

\correspondingauthor{Marta Frias Castillo}
\email{friascastillom@strw.leidenuniv.nl}

\author[0000-0002-9278-7028]{Marta Frias Castillo}
\affiliation{Leiden Observatory, Leiden University, Niels Bohrweg 2, 2333 CA Leiden, the Netherlands}

\author[0000-0002-1383-0746]{Matus Rybak}
\affiliation{Leiden Observatory, Leiden University, Niels Bohrweg 2, 2333 CA Leiden, the Netherlands}
\affiliation{Faculty of Electrical Engineering, Mathematics and Computer Science, Delft University of Technology, Delft, The Netherlands}

\author[0000-0001-6586-8845]{Jacqueline Hodge}
\affiliation{Leiden Observatory, Leiden University, Niels Bohrweg 2, 2333 CA Leiden, the Netherlands}

\author[0000-0001-5434-5942]{Paul van der Werf}
\affiliation{Leiden Observatory, Leiden University, Niels Bohrweg 2, 2333 CA Leiden, the Netherlands}

\author[0000-0001-9585-1462]{Dominik A. Riechers}
\affiliation{I. Physikalisches Institut, Universit\"at zu K\"oln, Z\"ulpicher Strasse 77, D-50937 K\"oln, Germany}

\author{Daniel Vieira}
\affiliation{Department of Astronomy, Cornell University, Space Sciences Building, Ithaca, NY 14853, USA}
\affiliation{I. Physikalisches Institut, Universit\"at zu K\"oln, Z\"ulpicher Strasse 77, D-50937 K\"oln, Germany}

\author{Gabriela Calistro Rivera}
\affiliation{European Southern Observatory (ESO), Karl-Schwarzschild-Stra\ss e 2, 85748 Garching bei M\"unchen, Germany }

\author{Laura N. Mart\'inez-Ram\'irez}
\affiliation{European Southern Observatory (ESO), Karl-Schwarzschild-Stra\ss e 2, 85748 Garching bei M\"unchen, Germany }
\affiliation{Instituto de Astrof\'isica, Facultad de F\'isica, Pontificia Universidad Cat\'olica de Chile Av. Vicuna Mackenna 4860, 782-0436 Macul, Santiago, Chile}
\affiliation{Escuela de F\'isica - Universidad Industrial de Santander, 680002 Bucaramanga, Colombia}

\author[0000-0003-4793-7880]{Fabian Walter}
\affiliation{Max Planck Institute for Astronomy, K\"onigstuhl 17, 69117,  Heidelberg, Germany}
\affiliation{National Radio Astronomy Observatory, Pete V. Domenici Array Science Center, P.O. Box O, Socorro, NM 87801, USA}

\author[0000-0001-8957-4518]{Erwin de Blok}
\affiliation{Netherlands Institute for Radio Astronomy (ASTRON), Postbus 2, 7990 AA Dwingeloo, the Netherlands}
\affiliation{Dept.\ of Astronomy, Univ.\ of Cape Town, Private Bag X3, Rondebosch 7701, South Africa}
\affiliation{Kapteyn Astronomical Institute, University of Groningen, Postbus 800, 9700 AV Groningen, The Netherlands}

\author[0000-0002-7064-4309]{Desika Narayanan}
\affiliation{Department of Astronomy, University of Florida, 211 Bryant Space Sciences Center, Gainesville, FL 32611 USA}
\affiliation{University of Florida Informatics Institute, 432 Newell Drive, CISE Bldg E251, Gainesville, FL 32611}
\affiliation{Cosmic Dawn Center at the Niels Bohr Institute, University of Copenhagen and DTU-Space, Technical University of Denmark}

\author{Jeff Wagg}
\affiliation{SKA Observatory, Lower Withington Maccleseld, Cheshire SK11 9DL, UK}

\begin{abstract}
We present a high-resolution study of the cold molecular gas as traced by CO(1-0) in the unlensed z$\sim$3.4 submillimeter galaxy SMM J13120+4242, using multi-configuration observations with the Karl G. Jansky Very Large Array (JVLA). The gas reservoir, imaged on 0.39$"$ ($\sim$3 kpc) scales, is resolved into two components separated by $\sim$11 kpc with a total extent of 16$\,\pm\,$3 kpc. Despite the large spatial extent of the reservoir, the observations show a
CO(1-0) FWHM linewidth of only 267$\,\pm\,$64 km s$^{-1}$. We derive a revised line luminosity of L$'_{\text{CO(1-0)}}\,=\,$(10 $\pm$ 3)$\,\times$ 10$^{10}\,$K$\,$km$\,$s$^{-1}\,$pc$^2$ and a molecular gas mass of M$_{\text{gas}} =\,$(13 $\pm$ 3)$\,\times\,$10$^{10}$ ($\alpha_\mathrm{CO}/1)$ M$_\odot$. Despite the presence of a velocity gradient (consistent with previous resolved CO(6-5) imaging), the CO(1-0) imaging shows evidence for significant turbulent motions which are preventing the gas from fully settling into a disk. The system likely represents a merger in an advanced stage. Although the dynamical mass is highly uncertain, we use it to place an upper limit on the CO–to–H$_2$ mass conversion factor $\alpha_\mathrm{CO}$ of 1.4. We revisit the SED fitting, finding that this galaxy lies on the very massive end of the main sequence at z$=$3.4. Based on the low gas fraction, short gas depletion time and evidence for a central AGN, we propose that SMM J13120 is in a rapid transitional phase between a merger-driven starburst and an unobscured quasar. The case of SMM J13120 highlights the how mergers may drive important physical changes in galaxies without pushing them off the main sequence.
\end{abstract}


\section{Introduction}
The evolution of the cosmic star formation rate density (SFRD) has now been well characterised as far back as the epoch of reionisation. It increased from the early stages of the Universe until it peaked at z $= 1 - 3$, and then decreased until the present day \citep{madau_dickinson2014}. Parallel to these optical and near-IR studies, powerful radio and submillimeter interferometers such as the Karl G. Jansky Very large Array (VLA), the Atacama Large Millimeter/submillimeter Array (ALMA) or the NOrthern Extended Millimeter Array (NOEMA) have enabled the first deep blind emission line surveys (ASPECS, at z$\sim 1 - 4$, \citealt{walter2016}, 
COLDz, at z $\sim 2.5 - 6$, \citealt{pavesi2018}). These have begun to unveil the evolution of the cosmic molecular gas mass density out to z $\sim$ 6, revealing a very good agreement with the cosmic SFRD. Along with progress in state-of-the-art hydrodynamical simulations (e.g., SIMBA, \citealt{dave2019}; IllustrisTNG, \citealt{diemer2019,popping2019illustris}),
these studies add to the growing consensus that the molecular gas reservoirs of galaxies are key drivers of the observed cosmic SFRD evolution, rather than changes in star forming efficiency \citep{scoville2016,tacconi2018,riechers2019,decarli2020,walter2020}. Understanding the molecular gas properties of galaxies over cosmic time is therefore crucial for obtaining a complete picture of how galaxies form and evolve \citep{kennicutt+evans2012,carilli-walter2013,hodge+dacunha2020,tacconi2020}. 

$^{12}$CO (hereafter CO), the second most abundant molecule in galaxies after molecular hydrogen (H$_2$), has been traditionally used to trace the total molecular gas content of galaxies. The low excitation temperature T$_{ex}$ = 5.5 K of the rotational ground state ($J = 1 - 0$) of CO means that this molecule is easily excited in a variety of galactic environments, tracing the bulk of the cold molecular gas. The CO(1-0) luminosity can be directly related to the cold molecular gas mass via a conversion factor $\alpha_\mathrm{CO}$ \citep[see][for a review]{bolatto2013}. Due to the intrinsic faintness of the line and the great distances involved, large time investments are needed to detect this transition at high redshift, so mid- and high-J CO transitions such as $J = 4 - 3$ and $J = 6 - 5$ 
are often used instead. These transitions trace denser, actively star-forming gas, and are typically brighter and thus easier to detect than the ground state transition. Relating their line luminosities to the CO(1-0) line luminosity, however, requires additional assumptions about line excitation, with conversion factors spanning a wide range of values over the high-z galaxy population  \citep{carilli-walter2013,narayanan_krumholz2014,sharon2016,yang2017,riechers2020a-vlaspecs}. Furthermore, higher-J CO transitions have been shown to miss a significant fraction of the lower-excitation gas, and might therefore not be representative of the total molecular gas content of the galaxies. This would in turn lead to an underestimation of the dynamical mass and gas fraction, among other properties \citep{ivison2011,riechers2011,Canameras2018}.

Submillimeter galaxies (SMGs) are one of the main sites of star formation in the high-z Universe \citep{casey2014,hodge+dacunha2020}. These dusty, high infrared luminosity (L$_\mathrm{IR} >$ 10$^{12}$ L$_\odot$; \citealt{magnelli2012lir_smg}) systems are found predominantly at high redshift \citep{chapman2005,danielson2017}. They have star formation rates up to, and even in excess of, $\sim$ 1000 M$_\odot$ yr$^{-1}$, fed by large molecular gas reservoirs of 10$^{10-11}$ M$_\odot$ \citep{greve2005,tacconi2008,bothwell2013,birkin2020}. Such intense starburst episodes are thought to be mainly triggered by major mergers, interactions with  neighbouring galaxies or through accretion of cold gas from the intergalactic medium. Obtaining a clear picture of the physical conditions and structure of their interstellar medium (ISM), as well as the processes regulating their star formation, is critical to our understanding of the history of cosmic star formation. However, despite their extreme properties, resolved studies of the low-J CO emission in high-redshift SMGs on kpc/sub-kpc scales remain a technical challenge, even for the few extremely bright \citep[e.g., GN20,][]{hodge2012} and/or strongly lensed systems (e.g., HLSJ091828.6+514223; \citealt{rawle2014}; SDSS J0901+1814; \citealt{sharon2019}) in which they have been carried out. In this paper we extend these to SMM J13120+4242 (hereafter SMM J13120).

SMM J13120 is a z $= 3.4$, non-lensed SMG that was first detected in the Hawaii Deep Field SSA 13. This SMG has an active galactic nucleus (AGN), revealed by optical emission lines \citep[Si IV, C IV, O III,][]{chapman2005}, point-like X-ray imaging \citep{J13120chandra} and the need to include an AGN component to accurately fit the SED (Section \ref{sec:results}). CO(1-0) observations with the GBT by \cite{hainline2006} originally suggested a very broad line of FWHM $\sim$ 1000 km s$^{-1}$, implying an ongoing merger. This scenario was further supported by resolved
observations of CO(6-5) showing a disturbed structure and velocity distribution \citep{engel2010}. Subsequent CO(1-0)
imaging with the VLA revealed a very massive (M$_\mathrm{gas}$ = 1.9-6.9$\times$10$^{11}$ M$_\odot$), extended (15 kpc) low excitation gas reservoir (\citealt{riechers2011}, hereafter R11). In this work we present high-resolution (0.39$"$) JVLA CO(1-0) imaging of J13120, spatially and kinematically resolving the cold molecular gas reservoir in this unlensed galaxy. 

This paper is organised as follows. In Section \ref{sec:observations}, we describe the observations and data reduction. The main results are presented in Section \ref{sec:results}, where we include a reevaluation of the evidence for a broad CO(1-0) linewidth, source size estimation and revisit the SED fitting. In Section \ref{sec:analysis} we present the analysis of this data, including estimates of the gas fraction and gas depletion time (Section \ref{sec4.1}), the dynamical mass and CO-to-H$_2$ conversion factor (Section \ref{sec:DynamicalModel}) and the revised gas excitation modeling (Section \ref{lvg}). In Section \ref{sec:discussion} we place the extended gas reservoir seen in SMM J13120 in context with the literature (Section \ref{reservoir_context}) and we further discuss the fate of this galaxy given the analysis presented here (Section \ref{sec5.2}). We end with our conclusion in Section \ref{sec:conclusion}.

A cosmology of H$_0$= 67.4 km s$^{-1}$ Mpc$^{-1}$, $\Omega_M$ = 0.315, and $\Omega_\Lambda$ = 0.685 is assumed throughout this paper \citep{planck2020A&A...641A...6P}. At a redshift of z = 3.408, this corresponds to an angular scale of 7.56 kpc arcsec$^{-1}$ and a luminosity distance of 30.3 Gpc.

\section{Observations and data reduction}
\label{sec:observations}
\subsection{JVLA K-Band}
We used the JVLA to observe the CO(J = 1-0) emission (rest-frame frequency: $\nu_\mathrm{rest}$ = 115.27 GHz) from SMM J13120 (VLA program ID: 15A-405 and 17B-108, PI: Hodge). At z = 3.408, the line is redshifted to 26.15 GHz.
We used the K band receivers in combination with the WIDAR correlator configured to 8 bit sampling mode to observe a contiguous bandwidth of 2.048 GHz (full polarization) covering the 24.446 - 26.342 GHz frequency range at 2 MHz spectral resolution (23 km s$^{-1}$ at 26.15 GHz). Four of the allocated 50 hours were observed in 2015 March (B configuration). The remaining 46 hrs were observed between 2018 October and December (B and C configurations). In total, 39.9 hours were spent in B configuration and 9.96 hours in C configuration. The data were combined with previous observations taken between 2010 May (C and D configurations) and 2011 January (CnB configuration) (Program ID: AR708, PI: Riechers). This resulted in a combined integration time of 57.6 hrs. After accounting for calibration overheads and flagging, the total on-source integration time was 30.9 hrs.

The quasar 3C 286 was used to determine the absolute flux density and for bandpass calibration. The VLA calibrator J1242+3751 was observed every 3.5 minutes for phase and amplitude calibration. Observations were taken in full polarization mode.

All data was reduced using the Common Astronomy Software Application (CASA) package version 5.6.0-60 \citep{casa}. Several antennas and two spectral windows were manually flagged after inspecting the calibration output. In K-band, the largest angular scale is 7.9" (59.5 kpc at z$\sim$3.4) for B configuration and 66" (497.6 kpc) for C and D configurations at 22GHz, the center of the band.

We imaged the data using the CLEAN algorithm, and cleaned down to 2$\sigma$ in a 5.0" $\times$ 4.3" clean box around SMM J13120+4242. We used natural weighting, which resulted in a synthesised beam FWHM of 0.39" $\times$ 0.34" at a PA = -81$\degree$. We further performed synthesis imaging of the data using the multi-scale version of the WS-Clean algorithm introduced by \citet{offringa2014,offringa2017}, but found no quantitative or qualitative improvement over the imaging performed with CASA. The resulting rms value achieved was 27.2 $\mu$Jy beam$^{-1}$ channel$^{-1}$, with channels of width 46 km s$^{-1}$. At a tapered resolution of 0.9$"$, we achieve an rms of 35 $\mu$Jy beam$^{-1}$ channel$^{-1}$, in comparison to the 70 $\mu$Jy beam$^{-1}$ channel$^{-1}$ depth of \citetalias{riechers2011} for an equal channel width and spatial resolution.

\subsection{Plateau de Bure Interferometer}
Observations of the CO J = 6-5 transition were carried out in winter 2007/2008 with the IRAM Plateau de Bure interferometer in the extended A configuration. At z=3.4, the CO(6-5) transition is redshifted into the 2mm band. The data were presented in \citet{engel2010}. We reuse the imaging products created in the MAPPING environment of GILDAS (Engel, Davies, priv.comm.). The rms value is 0.64 mJy beam$^{-1}$ channel$^{-1}$ (channel width of 20 km s$^{-1}$), with a beam size of 0.59$"$ $\times$ 0.47$"$ at a P.A. = 50.9$\degree$.

\medskip
\section{Results} \label{sec:results}

\medskip
\subsection{CO(1-0) Line Detection}

\begin{figure}
    \hspace{-0.9cm}
    \includegraphics[scale=0.6]{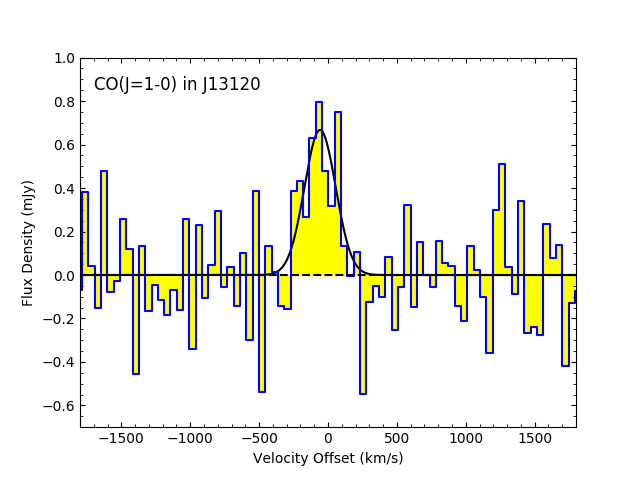}
    \caption{CO(1-0) spectrum of SMM J13120 with a spectral resolution of 46 km s$^{-1}$. The solid line shows the Gaussian fit to the data, which yields a peak flux density of 0.67 $\pm$ 0.14 mJy and a FWHM of 267 $\pm$ 64 km s$^{-1}$.}
    \label{line_profile}
\end{figure}

The CO(1-0) spectrum of SMM J13120 is shown in Fig. \ref{line_profile}. The spectrum was extracted using a circular aperture of 1.75$"$ diameter centered on the source, chosen to maximise the recovered flux. We fit the line with a single Gaussian profile that yields a peak flux density of 0.67 $\pm$ 0.14 mJy, with a FWHM of 267 $\pm$ 64 km s$^{-1}$ and integrated flux of 0.19 $\pm$ 0.06 Jy km s$^{-1}$. The line is consistent with the central peak that was detected at a 4$\sigma$ level in the earlier VLA observations by \citetalias{riechers2011} (see their Figure 2, bottom), which we use as our short spacings here. 

\begin{figure*}
    \hspace{-3cm}
    \includegraphics[scale=0.6]{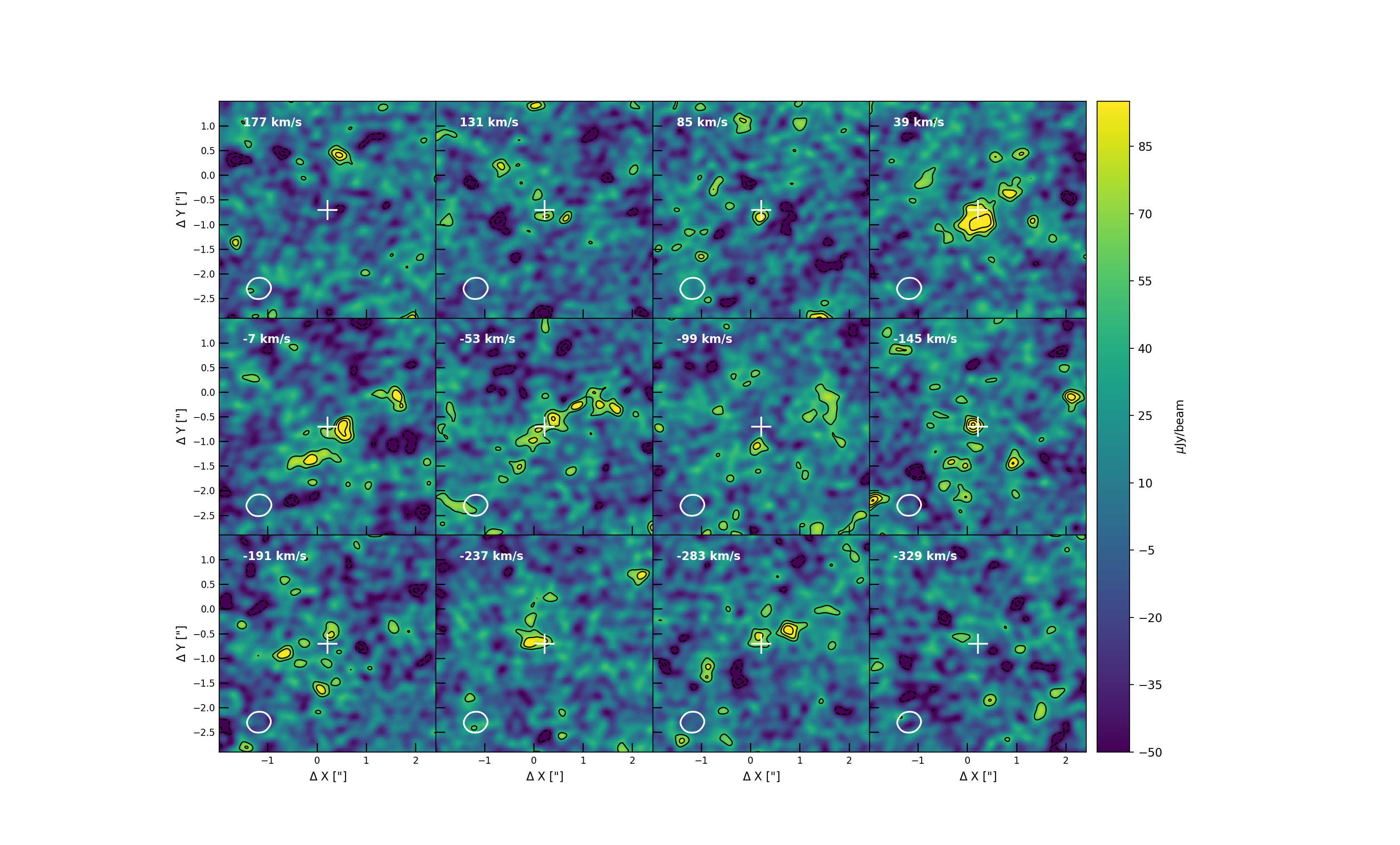}
    \caption{CO(1-0) emission maps in SMM J13120 in 46 km s$^{-1}$ channels. Cleaned maps are shown at native resolution of 0.39$"$ from 177 km/s in the top left channel to -329 km/s in the bottom right. The shape of the beam is displayed in each channel. The white cross corresponds to the best fit peak position of a 2D elliptical Gaussian to the 0th moment map. Contours start at $\pm$2$\sigma$ and increase in steps of $\sigma$ = 27.7 $\mu$Jy beam$^{-1}$.}
    \label{chans}
\end{figure*}

Channel maps of 46 km s$^{-1}$ width are shown in Figure \ref{chans}. Emission first appears at 131 km s$^{-1}$, and disappears at -283 km s$^{-1}$.
The data cube was tapered to 0.6$"$ and 0.9$"$ to search for additional extended emission that might be resolved out at native resolution (see Appendix), but no significant difference in total flux density was found. 

Ancillary data for SMM J13120 previously suggested a broader CO line FWHM. \citet{greve2005} detected unresolved CO(4--3) emission with PdBI with a velocity width of 530 $\pm$ 50 km s$^{-1}$. \citet{engel2010} reported a broad $\sim$900 km s$^{-1}$ FWHM for the CO(6-5) emission. Upon examination of the CO(6-5) data cube (\citealt{engel2010}, priv.\ comm.; see Figure \ref{pv_overplot}.), we found that the SNR of the data and the narrow total bandwidth of the cube (1800 km s$^{-1}$) does not allow a robust continuum subtraction, so it is possible that continuum emission is causing an artificial widening of this line. 

The earlier spectra of SMM J13120 from the GBT \citep[using the Autocorrelation Spectrometer;][]{hainline2006} and VLA \citepalias{riechers2011} appeared to show evidence for CO(1-0) emission as well across a broader, $>$1000 km s$^{-1}$ velocity range, albeit at limited SNR for both instruments. This naturally raises a question about whether the new high-resolution VLA data are missing more extended emission at large velocities, such as has been reported in some follow-up studies of high-redshift sources with the GBT \citep[e.g.,][]{frayer2018}. However, the new data disfavour the presence of such a broad component at a flux level consistent with the uncertainties on the earlier VLA data (see Appendix). Moreover, despite the two times improved sensitivity of the combined dataset even when tapering to the resolution of the initial data set, we do not find evidence for emission above 2$\sigma$ significance beyond the central emission component. We note that, despite the narrow linewidth we find here, we still find evidence for a spatially extended CO(1-0) reservoir (Section 3.2) and a diffuse low-excitation gas component (Section \ref{lvg}). A deep integration with a single-dish telescope would be required to investigate if an even lower surface brightness extended component is present in this source.

\begin{table}
\caption{Observed and derived properties of SMM J13120+4242. \label{tab:results} }
\begin{center}
 \begin{tabular}{@{}lc @{}}
 \hline \hline
Parameter & Value\\
\hline
Position (J2000) & 13h12m01.182s \ +42$\degree$42$'$08.276$"$\\
z$^a$ & 3.4078 $\pm$ 0.0014\\
S$_\mathrm{CO(1-0)}$ &  0.67 $\pm$ 0.14 mJy\\
FWHM$_\mathrm{CO(1-0)}$ & 267 $\pm$ 64 km s$^{-1}$\\
FWHM$_\mathrm{CO(6-5)}$$^b$ & 911 $\pm$ 140 km s$^{-1}$\\
I$_\mathrm{CO(1-0)}$ & 0.19 $\pm$ 0.06 Jy km s$^{-1}$\\
I$_\mathrm{CO(6-5)}$$^b$ & 2.54 $\pm$ 0.1 Jy km s$^{-1}$\\
L$'_\mathrm{CO(1-0)}$  & (10 $\pm$ 3) $\times$ 10$^{10}$ K km s$^{-1}$ pc$^2$\\
L$_\mathrm{CO(1-0)}$  &  (4.7 $\pm$ 1.5) $\times$ 10$^{6}$ L$_\odot$\\
R$_{1/2}^\mathrm{CO(1-0)}$ & 8 $\pm$ 2 kpc\\
M$_\mathrm{gas}$ & (13 $\pm$ 3) $\times$ 10$^{10}$ M$_\odot$\\
M$_\mathrm{dyn}$ & (1.6 $\pm$ 0.9) $\times$ 10$^{11}$ M$_\odot$\\
L$_\mathrm{FIR}$ & (6.4 $\pm$ 2.8) $\times$ 10$^{12}$ L$_\odot$\\
SFR$_\mathrm{IR}$ & 948$^{+420}_{-418}$  M$_\odot$/yr \\
M$_{*}$ & 6.45$^{+2.5}_{-1.9}$ $\times$ 10$^{11}$ M$_{\odot}$\\
\hline
  \hline
$^a$ \citet{riechers2011} \\
$^b$ \citet{engel2010}
 \end{tabular}
 
\end{center}

\end{table}

\medskip
\subsection{Moment Maps}

\begin{figure}
    \centering
    \includegraphics[width=\linewidth]{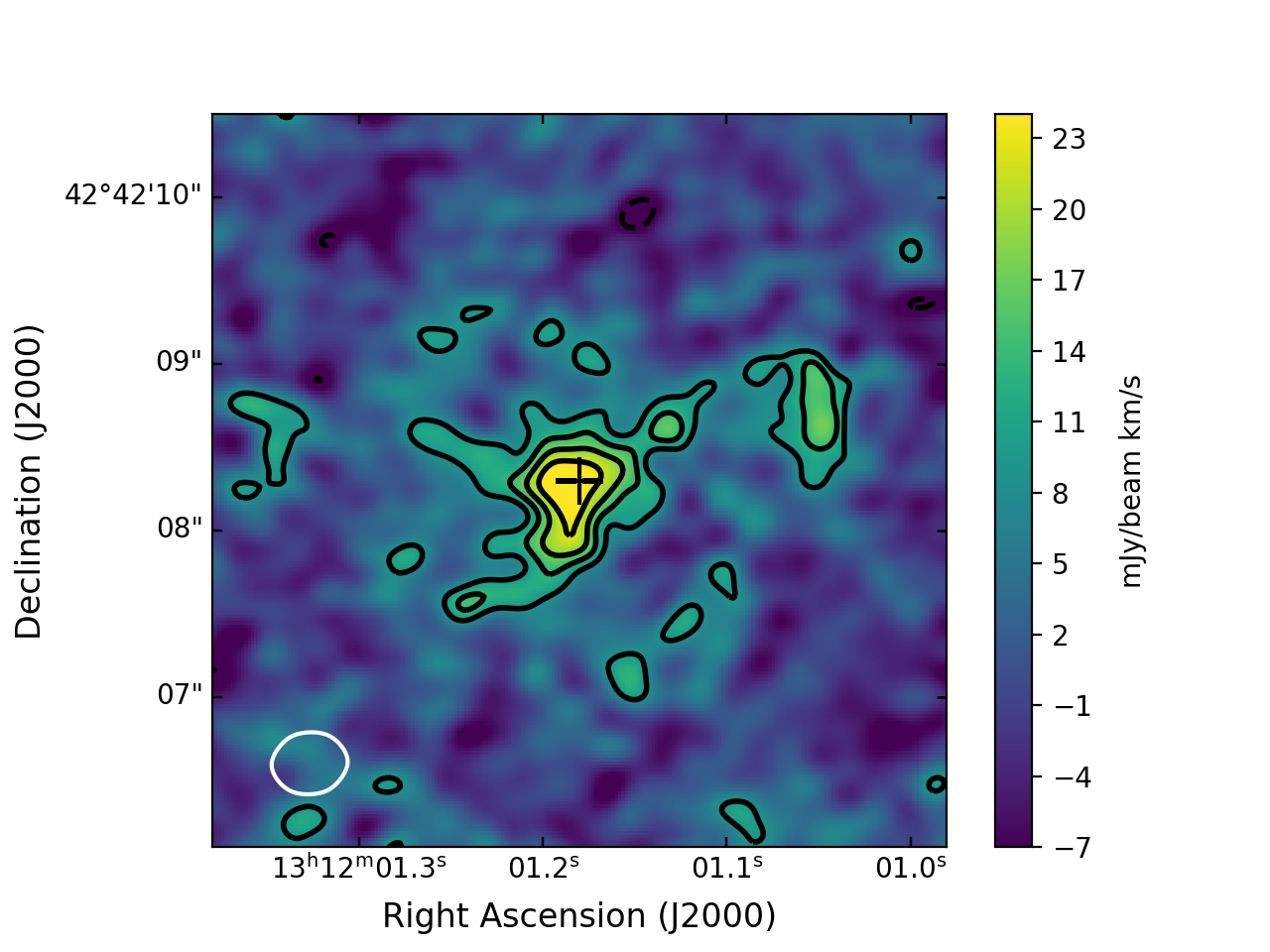}
    \includegraphics[width=\linewidth]{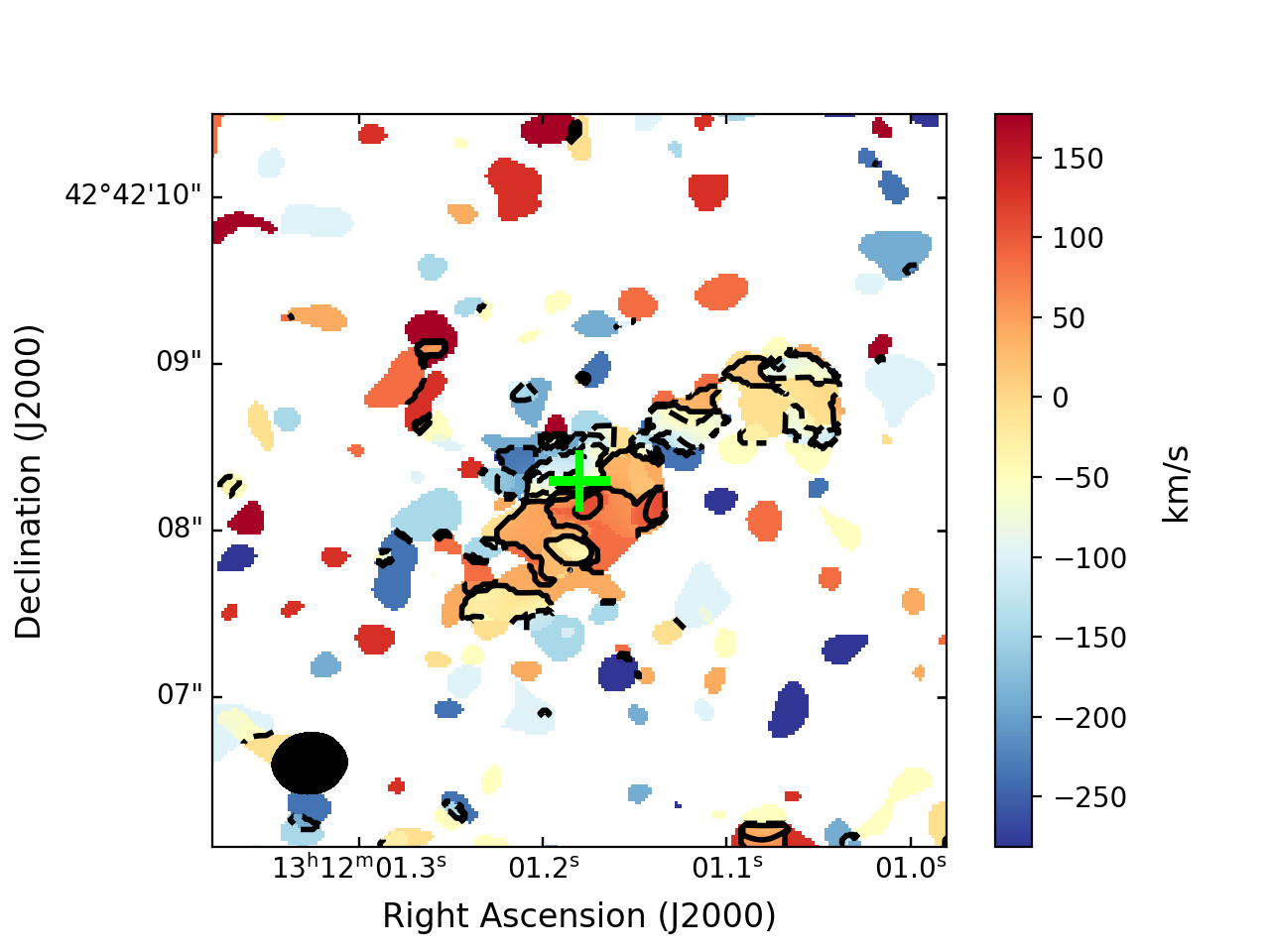}
    \caption{CO(1-0) 0th (top) and 1st (bottom) moment map of SMM J13120 at a resolution of 0.39$"$. The 0th moment map contours start at $\pm$ 2$\sigma$ and increase in steps of $\sigma$ = 4.4 mJy km s$^{-1}$. The 1st moment map was clipped at 2 times the rms noise per channel ($\sigma$ = 9 $\mu$Jy beam$^{-1}$), and contours are shown in steps of 50 km s$^{-1}$. We fit a 2D elliptical Gaussian to the 0th moment map to determine the peak position of the emission, shown by the cross, corresponding to the white cross in Fig. \ref{chans}.}
    \label{moment_maps}
\end{figure}

The task \texttt{IMMOMENTS} in CASA was used to obtain the integrated-intensity map (moment 0) and intensity-weighted velocity field (moment 1). Both moment maps were extracted by collapsing the data cube over the channels where emission is present, and the moment 1 map was further clipped at 2 times the rms noise per channel. The maps are shown in Figure \ref{moment_maps}. The moment 0 map reveals a source elongated in the SE to NW direction, at a PA of 127$\degree$. We detect two different components separated by 1.5$"$ ($\sim$11 kpc). The central component carries most of the CO emission, with a luminosity ratio of 5:1. The secondary (NW) component is located at $\Delta$RA = -0.125$"$ (-0.9kpc) and $\Delta$Dec = 0.4$"$ (3kpc) with respect to the main component. 

The moment 1 map shows a complex velocity field, with most of the emission arising at the systemic velocity. There is no apparent structure in the velocity field of the NW component. No significant emission is seen outside this velocity range, which is consistent with the line profile discussed below. We do not detect any rest-frame 2.6mm continuum emission from either component down to a 3$\sigma$ limit of 10.6 $\mu$Jy beam$^{-1}$.

\medskip
\subsection{Source Size Estimation}

We fit a 2D Gaussian to the 0th moment map using the task \texttt{IMFIT} in CASA, finding a deconvolved size of
2.1 $\pm$ 0.5$"$ $\times$ 1.0 $\pm$ 0.2$"$ (16 $\pm$ 3 $\times$ 8 $\pm$ 2 kpc). For the two spatially resolved components seen in the moment 1 map, we find a size of 1.3 $\pm$ 0.3$"$ $\times$ 0.7 $\pm$ 0.2$"$ (10 $\pm$ 2 $\times$ 5 $\pm$ 1 kpc) and 0.6 $\pm$ 0.2$"$ $\times$ 0.3 $\pm$ 0.2$"$ (5 $\pm$ 1 $\times$ 2 $\pm$ 2 kpc) for the central and the NW component respectively. 

\begin{figure}
    \hspace{-0.8cm}
    \includegraphics[scale=0.6]{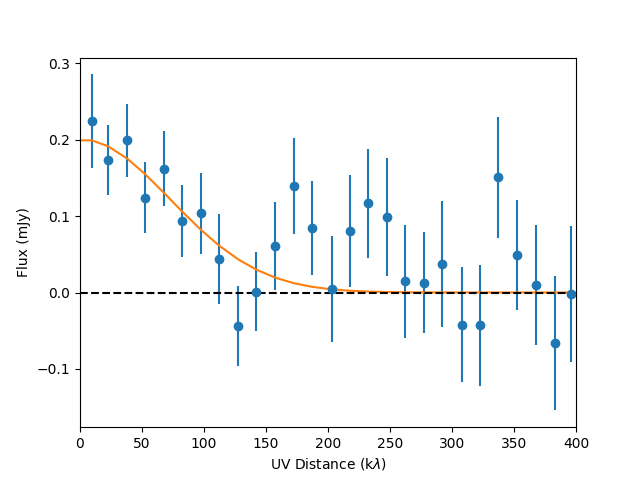}
    \caption{Visibilities (uv)-profile of J13120. The uv-data were extracted from the channels within the full-width at zero intensity of the line from the cube and averaged in bins of 15 k$\lambda$. We fit the profile with a single Gaussian to estimate the intrinsic source size, finding a half-light radius of 1.20 $\pm$ 0.15 arcseconds, consistent with the 2D elliptical Gaussian fit to the 0th moment map.}
    \label{uvfit}
\end{figure}

We further derived the size of the molecular gas reservoir from the uv-data directly to avoid possible biases introduced by the imaging process. We extract the visibilities corresponding to the channels within the full width at zero intensity of the line from the cube. We then average the data in radial bins of 15 k$\lambda$ and fit a Gaussian profile to the data, as shown in Figure \ref{uvfit}. The resulting half-light radius is 1.20 $\pm$ 0.15 $"$ (9.1 $\pm$ 1.1 kpc), consistent with our estimation from the moment 0 map fitting. At 150k$\lambda$ there is tentative evidence which may indicate the presence of unresolved structure not captured by our Gaussian model.

\medskip
\subsection{SED Fitting}
In order to perform a consistent analysis of SMM J13120, we re-examine the physical properties of the galaxy derived through SED fitting. Although previous estimates were reported in the literature \citep{hainline2006,michalowski2010}, we have included new photometric data available from the radio to the X-rays (see Table \ref{tab:ancillary}). In particular, a Chandra detection of this source \citep{J13120chandra} reported an X-ray luminosity of log(L$_\mathrm{0.5-8keV}$) $\sim$ 44, associated with an AGN (an unphysical SFR $\sim 10^5$ M$_{*}$ yr$^{-1}$ would be required to explain this emission from star formation, according to the L$_\mathrm{X-ray}$-SFR relation by \citealt{mineo2014}), which is furthermore supported by emission line properties characteristic of AGN \citep[]{chapman2005}.

\begin{figure*}
\centering
    \includegraphics[scale=0.6,angle=270]{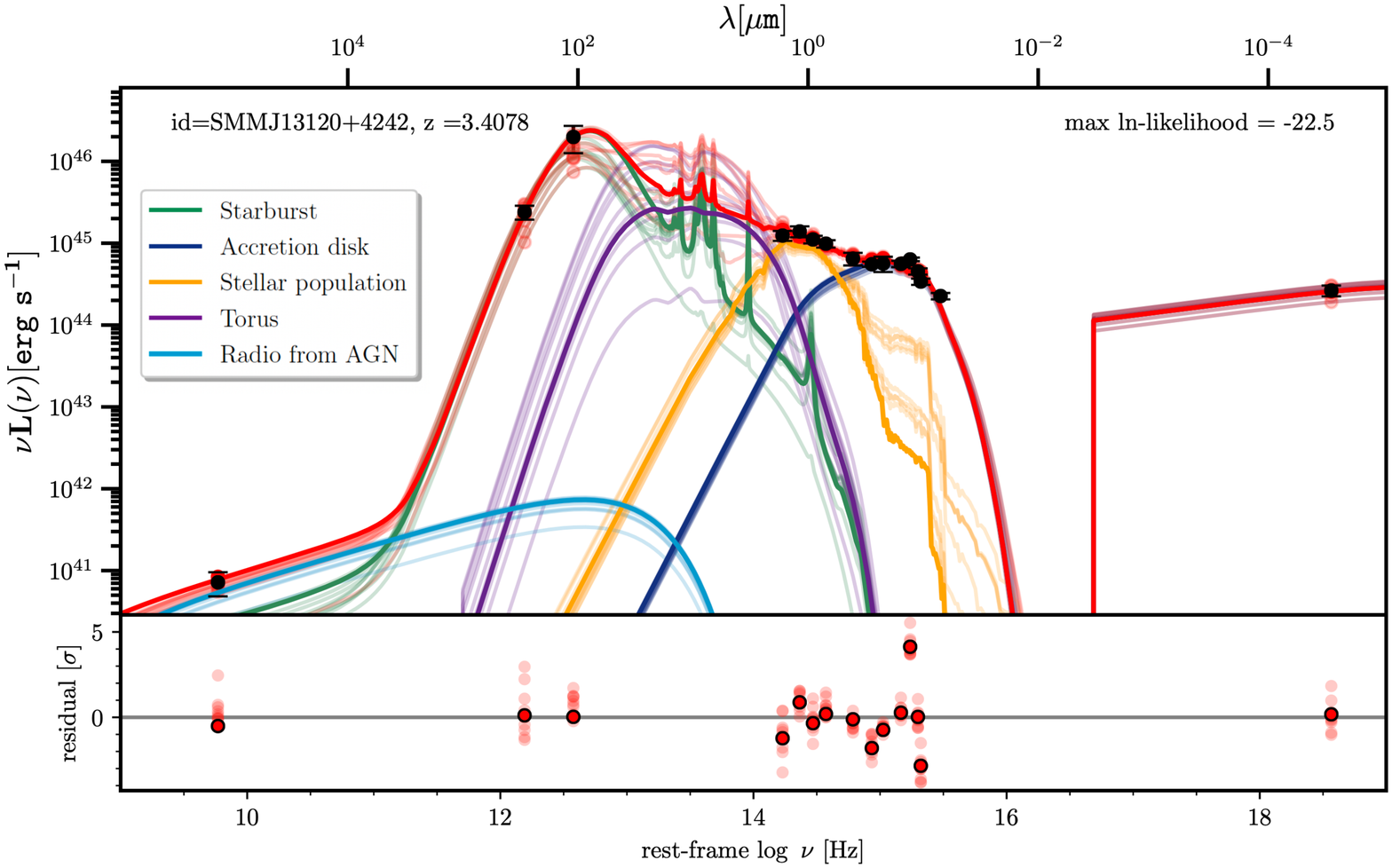}
    \caption{SED modelling of the radio-to-X-ray photometry available for J13120 using the Bayesian SED-fitting code AGNfitter-rX (Calistro-Rivera+16, Martinez-Ramirez, in prep.). In the main panel the photometric data points are shown with black error bars. The best fit and ten realisations of the posterior distributions of the fit are drawn as coloured solid and transparent lines, respectively. The realisations display the uncertainties associated with the fits. The different colours represent the total SED (red lines), the accretion disk and X-ray corona emission (dark blue), the host galaxy emission (yellow), the torus emission (purple), the galactic cold dust emission and radio emission that originates in star formation (green) and the radio AGN emission (turquoise). The lower panels show the residuals expressed in terms of significance given the data noise level for the best fit and the 10 realisations.}
    \label{SED}
\end{figure*}

To account for the contribution of the AGN to the multi-wavelength SED we use the most recent version of the MCMC-based code AGNfitter \citep{Calistro2016}. Given the unique data coverage for this source, we use the updated version of the code, AGNfitter-rX (Martinez-Ramirez in prep.), which is the first Bayesian SED-fitting routine which consistently models the SEDs of the host galaxy and the AGN emission, from the radio to the X-rays. The AGNfitter model we used here consists of six different physical components: the X-ray corona, the accretion disk emission \citep{richards2006}, the hot dust torus emission \citep{Stalevski2016}, the stellar populations \citep[][with a Chabrier IMF and exponentially declining star-formation history]{BC2003}, the cooler dust heated by star formation \citep[]{schreiber2016}, and the radio emission component (see Figure \ref{SED}). 
Our SED modelling approach yields a SFR$_\mathrm{IR}$=$948^{+420}_{-418}$ \rm  M$_{*}$ yr$^{-1}$ and a stellar mass of $\rm \log M_{*} [\mathrm{M_{\odot}}]=  11.81^{+0.15}_{-0.25}$. These estimates differ from those reported in the literature: \citet{hainline2008} found log(M$_{*}$) = 10.96±0.04 and SFR$_\mathrm{IR}$=890 M$_{*}$ yr$^{-1}$; whereas \citet{michalowski2010} found log(M$_{*}$)=11.36 (no uncertainties were reported) and 3700 M$_{*}$ yr$^{-1}$. The advantage of the new estimates lies on the improved modelling approach used, and the inclusion of the most recent additional data points in the fit (HST-WFC3 F160W and Chandra). Several tests were completed to assess the robustness of the derived values. Included within these were the use of restrictive and flexible energy balance (in which the luminosity of the cold dust is equal or allowed to be greater than the total luminosity attenuated by dust, respectively), setting the AGN component to 0 and deactivating all the priors. All the tests which allow an AGN component return stellar masses agreeing within the error bars. We note that the stellar masses inferred were not directly affected by the AGN component but rather by the additional photometric data points and different SFH models chosen. For the SFR$_\mathrm{IR}$, our estimate is consistent with the \citet{hainline2006} estimate, while it differs from \citet{michalowski2010}, which predicts an extremely large value that may be unrealistic given SMG statistics \citep{dacunha2015}. We note that a further significant advantage of the current method is its probabilistic Bayesian approach. The poor photometric coverage of the FIR-MIR SED directly translates into an uncertain SFR from the IR and an uncertain AGN torus contribution to the total IR emission (12\%$^{+34\%}_{-8\%}$ from our fit), therefore highlighting the need of a Bayesian approach that robustly recovers the uncertainties of the inferred quantities. With these uncertainties in mind, these revised estimates place SMM J13120 on the massive end of the main sequence at z = 3.4 (using the definition from \citealt{speagle2014}; see \citealt{hodge+dacunha2020} for a discussion of further systematic uncertainties).

\begin{table}
\caption{J13120 photometry. References are as follows: a) \citet[]{yang2017}, b) \citet[]{chapman2005}, c) \citet[]{fomalont2006}, d) \citet[]{smail2004}, e) \citet[]{swinbank2010}, f) \citet[]{hainline2008}, g) \citet[]{kovacs2006}. \label{tab:ancillary} }

 \begin{tabular}{@{}cc @{}}
 \hline 
Wavelength ($\mu$m) & Flux ($\mu$Jy) \\
\hline
(0.15 - 4) $\times$ 10$^{-3}$ $^a$  & 0.00029 $\pm$ 0.000043\\
0.428$^b$  & 0.302 $\pm$ 0.027\\
0.630$^c$  & 0.649 $\pm$ 0.057\\
0.656$^b$  & 0.912 $\pm$ 0.08 \\
0.767$^d$  & 1.4 $\pm$ 0.075\\
0.92$^c$  & 1.535 $\pm$ 0.135\\
1.25$^d$  & 2.138 $\pm$ 0.455\\
1.60$^e$ & 2.58 $\pm$ 0.19 \\
2.17$^d$ & 4.258 $\pm$ 0.749\\
3.6$^f$ & 10.6 $\pm$ 1.1\\
4.5$^f$ & 15.1 $\pm$ 1.6\\
5.8$^f$ & 24.1 $\pm$ 3.7\\
8.0$^f$ & 29.3 $\pm$ 4.1 \\
350$^g$  & 21100 $\pm$ 7700\\
850$^b$ & 6200 $\pm$ 1200 \\
214000 $^b$ & 49.1 $\pm$ 6 \\
\hline
\end{tabular}

\end{table}

\medskip
\section{Analysis} \label{sec:analysis}

\medskip
\subsection{Molecular Gas Mass, Gas Fraction and Depletion Time}\label{sec4.1}

The line flux was converted into line luminosity following \cite{solomon2005}:
\begin{equation}
    \mathrm{L'{_{CO}}} = 3.25 \times 10^7 \ \mathrm{I{_{CO}}} \ \mathrm{\nu{^{-2}_{obs}}} \ \mathrm{D{^2_{L}}} \ (1+\mathrm{z})^{-3} \ \mathrm{K \ km \  s^{-1} \ pc^2} ,
\end{equation}
\noindent where $\mathrm{I_{\text{CO}}}$ is the integrated line flux from the Gaussian fit in Jy km s$^{-1}$, $\mathrm{\nu_{\text{obs}}}$ is the observed frequency of the line and $\mathrm{D_{\text{L}}}$ is the luminosity distance in Mpc. 
We obtain a line luminosity of $\mathrm{L'_{CO(1-0)}}$ = (10 $\pm$ 3) $\times$ 10$^{10}$ K km s$^{-1}$ pc$^2$. This value falls within the scatter of the correlation between L$'_\mathrm{CO}$ and FWHM  found for high-redshift (z $\sim$ 1-6) SMGs both with direct CO(1-0) detections and based on high-J CO transitions \citep{yihan2019,birkin2020}.

The line luminosity can be transformed into a molecular gas mass through the conversion factor:
\begin{equation}
    \alpha_\mathrm{CO} = \frac{M_\mathrm{gas}}{L'_\mathrm{CO}} \ \mathrm{M_{\odot} \ K^{-1} \ km^{-1} \ s \ pc^{-2}}.
\end{equation}
It is common to apply different values for $\alpha_\mathrm{CO}$, depending on the galaxy type (see \citealt{bolatto2013} for a review). Normal star-forming galaxies show the standard Milky Way value $\alpha_\mathrm{CO}$ = 4.3 M$_{\odot}$ K$^{-1}$ km$^{-1}$ s pc$^{-2}$, while studies of local ULIRGS and high redshift starburst galaxies often assume $\alpha_\mathrm{CO}$ = 0.8 M$_{\odot}$ K$^{-1}$ km$^{-1}$ s pc$^{-2}$. Theoretical studies have shown that the value of $\alpha_\mathrm{CO}$ depends on a range of ISM conditions, such as metallicity and dust temperature \citep{narayanan2012, magnelli2012,gong2020}. 

There are a growing number of studies in high-z SMGs that point towards the use of $\alpha_\mathrm{CO} \sim$ 1 (e.g., \citealt{hodge2012,Rivera2018,birkin2020,riechers2021a}), so we adopt this value to estimate the molecular gas mass of SMM J13120 (as we show in Section \ref{sec:DynamicalModel}, our data excludes $\alpha_\mathrm{CO} >$ 1.4). Multiplying by $\times$1.36 to account for Helium, we obtain M$_\mathrm{gas}$ = (13 $\pm$ 3) $\times$ 10$^{10}$ ($\alpha_\mathrm{CO}$/1.0) M$_\odot$, a factor of two higher than the median value obtained for the latest compilation of SMGs in \citet{birkin2020} of M$_\mathrm{gas}$ = (6.7 $\pm$ 0.5) $\times$ 10$^{10}$ M$_\odot$. Our gas mass estimate is also several times higher than those of the ALESS sample \citep{swinbank2013} and the compilation of more than 700 SMGs selected from the AS2UDS survey \citep{Dudzeviciute2020}, although we point out that these were derived from dust masses assuming a fixed gas-to-dust ratio , $\delta_{gdr}$, of 90 and 100, respectively, instead of CO observations.

With our gas and stellar mass estimates, the resulting baryonic gas mass fraction is f$_{\text{gas}}$ = M$_{\text{gas}}$/(M$_{\text{gas}}$ + M$_{*}$) = 0.17 $\pm$ 0.03. This value is over a factor of two lower than the compilations of SMGs at z $\sim$ 2.8 - 3, which have a median f$_{\text{gas}} \sim$ 0.4 \citep{Dudzeviciute2020}, as well as other star forming galaxies at z $\sim$ 3.3 \citep{suzuki2020} and main sequence galaxies at z $\sim$ 2.5 \citep{saintonge2013,tacconi2018}.

Using our revised estimate of the total molecular gas reservoir and SFR, we can calculate the time that it would take for the system to deplete the current gas supply at its current SFR, t$_\mathrm{dep}$ = 137 $\pm$ 69 ($\alpha_\mathrm{CO}$/1) Myr. This indicates a phase of rapid star formation seen also in some other high-z SMGs and starburst galaxies \citep{tacconi2008,hodge2013,riechers2013,aravena2016,oteo2016,nayyeri2017}, or local ULIRGs \citep{solomon1997,combes2013}, and is  shorter than the estimates for the average SMG population at this redshift (e.g., \citealt{birkin2020,Dudzeviciute2020}) or predictions from simulations \citep{mcalpine2019}, which favour depletion timescales of the order of several hundred million years. We note here that, due to its large M$_*$ and SFR, SMM J13120 is a rare, extremely massive galaxy that will likely not be captured in most hydrodynamical simulations.

\medskip
\subsection{Dynamical Modeling and CO-H$_2$ Conversion Factor}\label{sec:DynamicalModel}
\begin{figure}
    \hspace{-0.5cm}
    \includegraphics[scale=0.45]{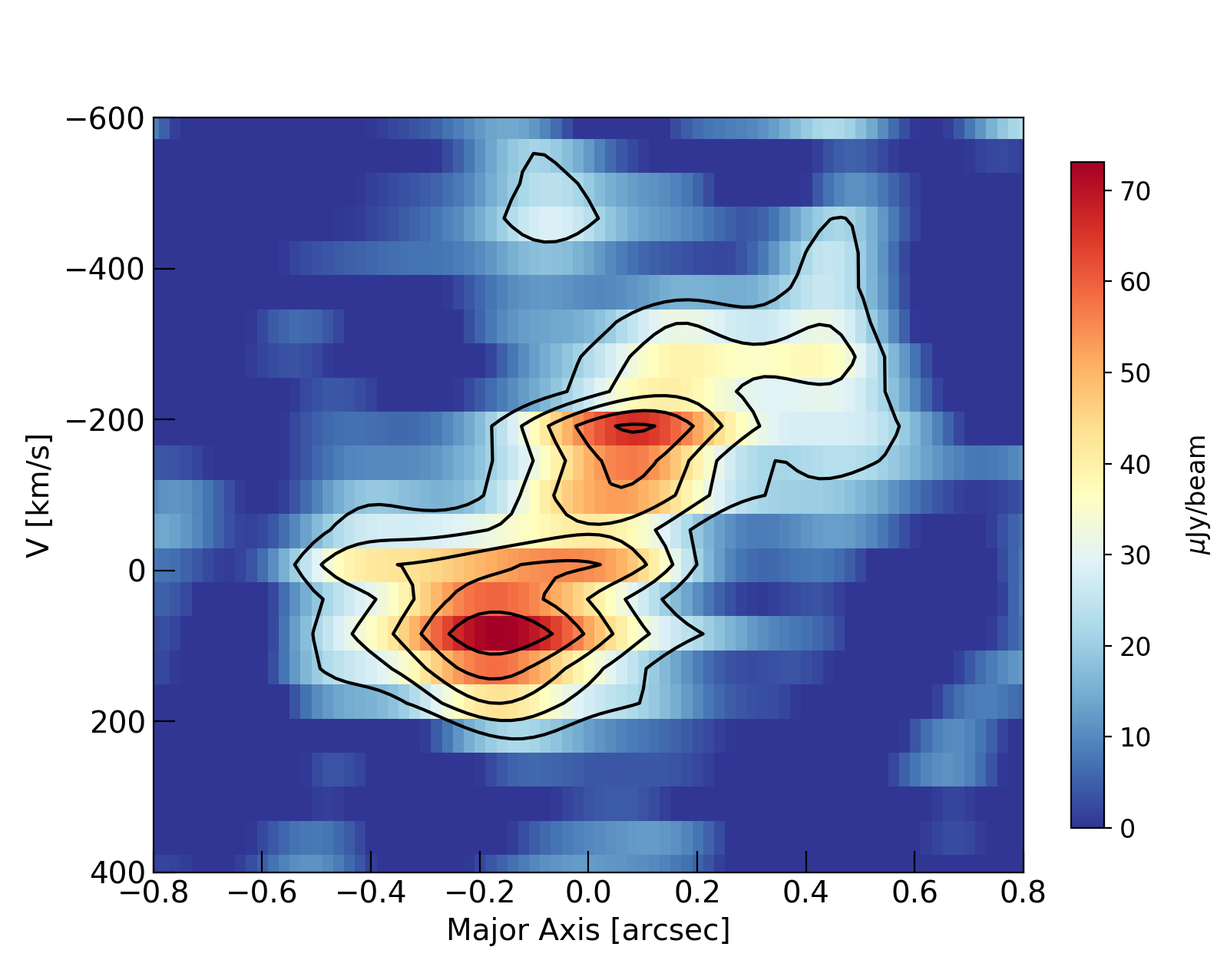}
    \caption{Position-velocity (PV) diagram of CO(1-0) emission in J13120. The axis of extraction is at PA of 42.7$\degree$ and an aperture of 0.3$"$ was used.
    Despite the disturbed velocity field seen in Fig. \ref{moment_maps}, bottom, there is a clear velocity gradient in SMM J13120.}
    \label{pv-diagram}
\end{figure}

\begin{figure}
    \hspace{-0.5cm}
    \includegraphics[height=7.5cm,width=9.1cm]{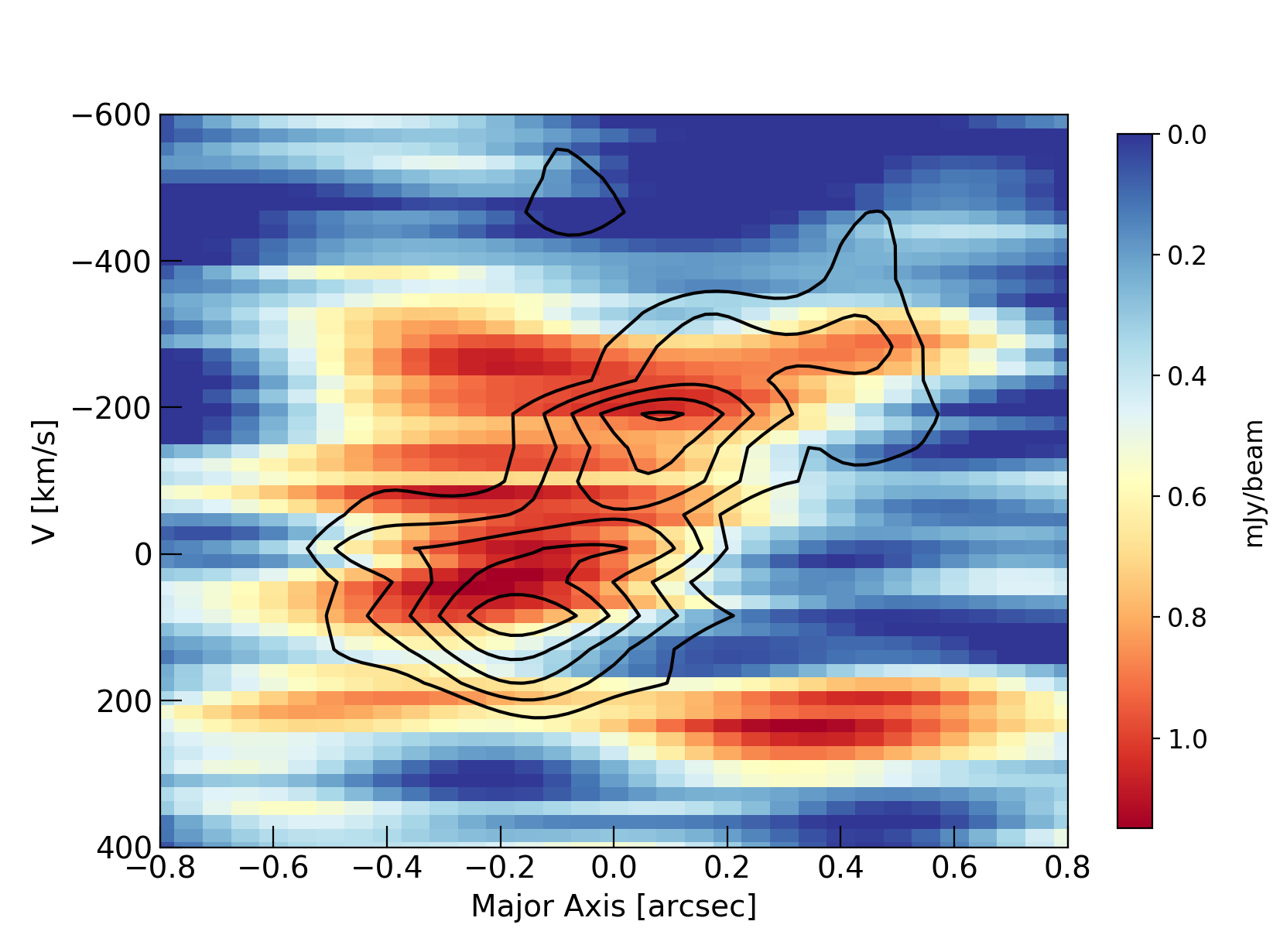}
    \caption{PV diagram of CO(6-5) emission in J13120 extracted from an axis with PA = 42.7$\degree$. Black contours correspond to the CO(1-0) PV diagram extracted from the same axis (see Figure \ref{pv-diagram}). The CO(6-5) emission shows evidence for a similar gradient as seen in CO(1-0).}
    \label{pv_overplot}
\end{figure}

We extracted position-velocity (PV) diagrams of the CO(1-0) and CO(6-5) emission along an axis at PA = 47.3\degree, chosen to maximise the observed velocity gradient. For comparison, the minor axis of the source has a PA = 37$\degree$. The resulting CO(1-0) PV diagram is shown in Figure \ref{pv-diagram}. A velocity gradient can be observed aligned with the minor axis of SMM J13120.  
In Fig. \ref{pv_overplot}  we also show the PV diagram extracted from the CO(6-5) cube along the same axis as before. Despite the perturbed structure of the velocity field seen in the CO(1-0) 1st moment map (Figure \ref{moment_maps}, bottom), a remarkably similar velocity gradient is present in both line transitions.  

We attempted to model the kinematics of SMM J13120 using the Bayesian Markov Chain Monte Carlo tools GaLPaK$^\mathrm{3D}$ \citep{galpak} and qubefit \citep{qubefit}. As noted above, however, the velocity gradient is aligned with the minor axis rather than the major axis. Together with the disturbed global morphology and limited SNR per channel, this precluded dynamically modelling the data as a rotating disk.

Instead we estimate the enclosed dynamical mass of SMM J13120 within the half-light radius through the measured CO(1-0) linewidth. We adopt the virial estimator \citep{tacconi2008,engel2010}, which relates the linewidth, radii and mass of the system through the equation:

\begin{equation}
M_\mathrm{dyn} (< r_{1/2}) = 2.8 \times 10^5 \ \Delta \text{v}^2 \ \mathrm{R_{1/2}} \ \mathrm{M_\odot},
\end{equation}

\noindent where $\Delta \text{v}$ is the line FWHM in km s$^{-1}$ and R$_{1/2}$ is the gas half-light radius in kpc. The scaling factor adopted assumes the gas to be dispersion dominated and is consistent with results derived from detailed Jeans modeling for high-z galaxies \citep{cappellari2009}. We consider this assumption to be appropriate given the chaotic velocity field in SMM J13120. We note that the scaling factor for a rotating disk at an average inclination would be a factor of $\sim$ 1.5 smaller. Taking the half-light radius as half of the semi-major axis FWHM (8 kpc, see Section \ref{sec:results}), we derive a value of M$_{\text{dyn}}$ = (1.6 $\pm$ 0.9) $\times$ 10$^{11}$ M$_\odot$.
This value is hard to reconcile with the stellar and gas mass estimates (6.45 $\times$ 10$^{11}$ M$_\odot$ and 1.3 $\times$ 10$^{11}$ M$_\odot$, respectively), highlighting the highly uncertain nature of the dynamical mass estimate due to the clear presence of non-gravitationally supported motions. Evidence of this comes from our inability to dynamically model the system, which suggests the CO(1-0) is dominated by turbulent motions which prevent the gas from settling into a disk. Nevertheless, we can attempt to put constraints on the $\alpha_\mathrm{CO}$ conversion factor by taking the 2$\sigma$ values of the quantities considered and assuming no dark matter. Taking the 2$\sigma$ values of M$_{\text{dyn}}$ and M$_{*}$, 3.4$\times$10$^{11}$ and 1.5$\times$10$^{11}$ respectively, we find $\alpha_\mathrm{CO}$ $<$ 1.4. If we instead consider M$_{\text{dyn}}$ = 2.5$\times$10$^{11}$ (+1$\sigma$), we find $\alpha_\mathrm{CO}$ $<$ 0.7. Even when making these liberal assumptions, we find that $\alpha_\mathrm{CO}$ cannot be higher than 1.4, consistent with recent work on SMGs at high redshift \citep{Rivera2018,birkin2020}. This supports the choice of a low, ULIRG-like $\alpha_\mathrm{CO}$ value as discussed in Section \ref{sec4.1}.

\medskip
\subsection{Gas Excitation}\label{lvg}

\begin{figure}
    \centering
    \includegraphics[width=\linewidth]{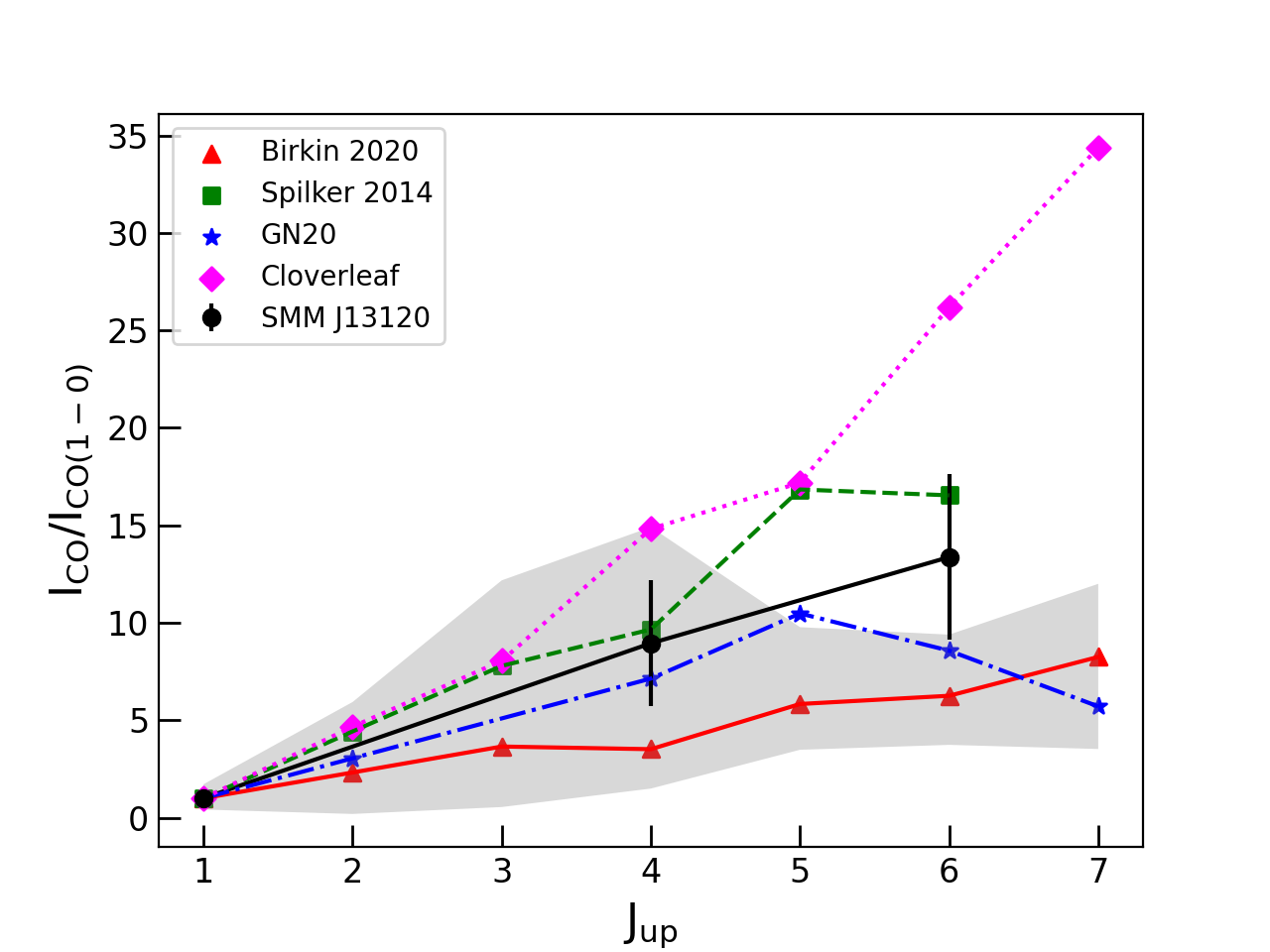}
    \includegraphics[width=\linewidth]{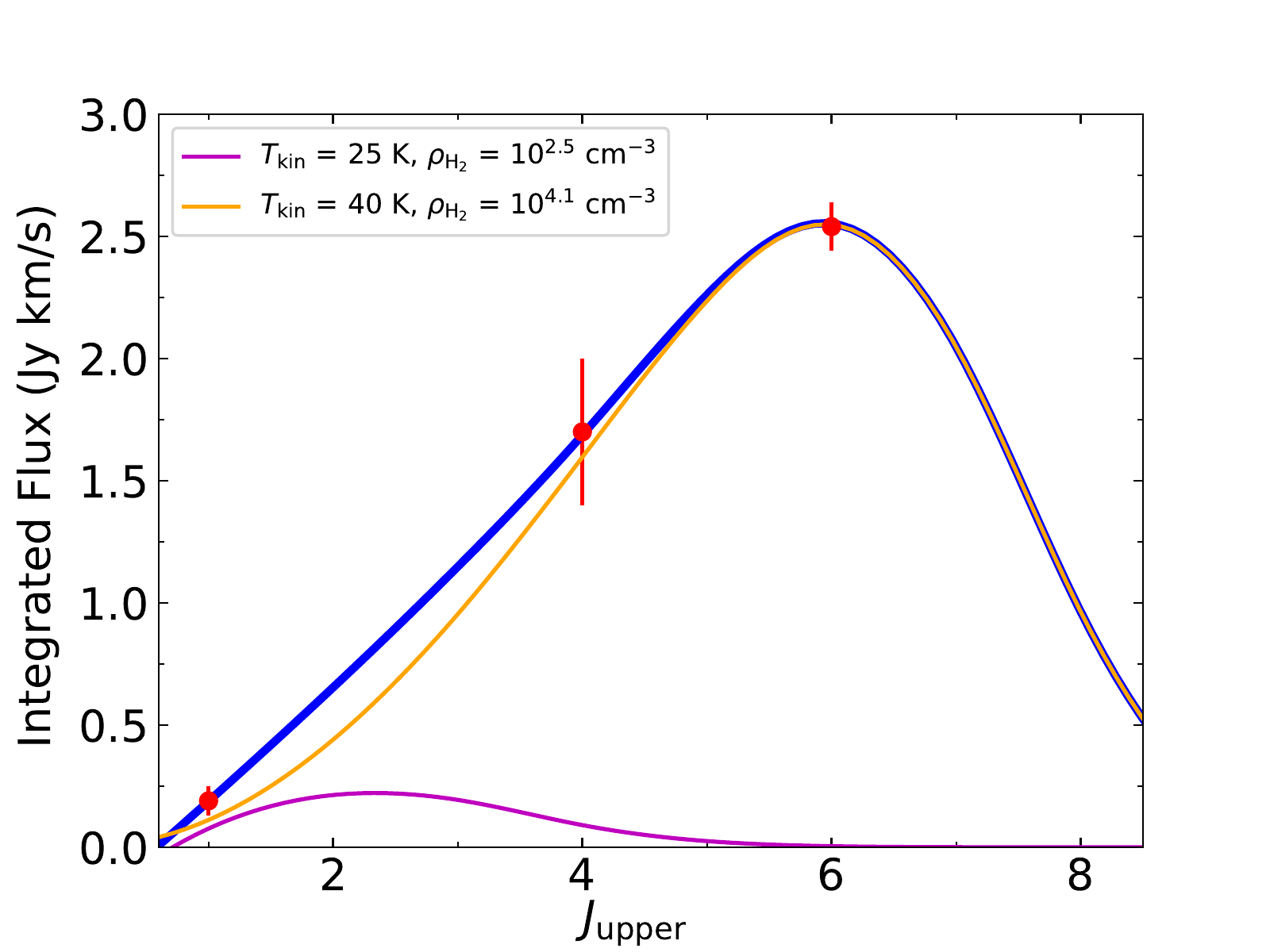}
    \caption{Top: Observed CO SLEDs normalised by the CO(1-0) flux for SMM J13120 (black circles), the SMG sample from \citet[red triangles]{birkin2020} and the SPT sample from \citet[green squares]{spilker2014}. The grey shadowed area represents the scatter of the \citet{birkin2020} sample, which decreases at higher-J due to the limited number of data points. Also shown for comparison are the SMG GN20 \citep[blue stars,][]{hodge2012} and the QSO Cloverleaf \citep[magenta diamonds,][]{bradford2009}. Bottom: CO excitation ladder (data points) and LVG model (solid line) for SMM J13120. The CO(J=4-3) and CO(6-5) data points are taken from \citet{greve2005} and \citet{engel2010}, respectively. The best fit to the data is a two-component LVG model (blue line) represented by a diffuse, low-excitation component with a kinetic temperature of T$_\mathrm{kin}$ = 25$\pm$5 K and a gas density of $\mathrm{\rho_{H_2}}$ = 10$^{2.5\pm0.1}$ cm$^{-3}$ (magenta line), and a high-excitation component with T$_\mathrm{kin}$ = 40$\pm$5 K and $\mathrm{\rho_{H_2}}$ = 10$^{4.1\pm 0.3}$ cm$^{-3}$ (yellow line).}
    \label{sled}
\end{figure}

We can build the CO SLED of SMM J13120 using the CO(4-3) and CO(6-5) line luminosities \citep{greve2005,engel2010}. We obtain line luminosity ratios of r$_{43/10}$ = 0.56 $\pm$ 0.20 and r$_{65/10}$ = 0.37 $\pm$ 0.12. Fig. \ref{sled}, top, shows the CO SLED in comparison with high-redshift systems GN20 \citep{carilli2010gn20sled} and the QSO Cloverleaf \citep{bradford2009}. We also show the statistical SLEDs derived for high-z SMGs in \citet{birkin2020} and for 1.4 mm-selected dusty star-forming galaxies in \citet{spilker2014}. The excitation in SMM J13120 agrees well with the general SMG population up to J = 4, with r$_{43/10,\mathrm{SMG}}$ = 0.32 $\pm$ 0.05, and displays a slightly higher excitation at J = 6, although still consistent with the average r$_{65/10,\mathrm{SMG}}$ = 0.3 $\pm$ 0.09 \citep{birkin2020}. 

The shape of the CO SLED can give us information on the gas density and temperature distributions. Therefore, we redo the Large Velocity Gradient (LVG) modeling of \citetalias{riechers2011} with the new CO(1-0) line flux. We calculate the brightness temperatures for the observed CO transitions on a grid of kinetic temperature and H$_2$ density spanning 10-200 K and 10$^{2.6}$ - 10$^{6.1}$ cm$^{-3}$. We fix the H$_2$ ortho-to-para ratio at 3:1 and set the CMB temperature to 12.03 K. We adopt a CO-H$_2$ abundance ratio per velocity gradient of 1 $\times$ 10$^{-5}$ pc (km s$^{-1}$)$^{-1}$ and use the \cite{flower2001} CO collision rates.
We then convert the Rayleigh-Jeans temperature to line intensities and run an MCMC code to obtain the optimal values with uncertainties for the gas kinetic temperature, H$_2$ density and source filling factor. We find that the data is well fit by a two-component LVG model, represented by a diffuse, low-excitation component with a kinetic temperature of T$_\mathrm{kin}$ = 25$\pm$5 K and a gas density of $\mathrm{\rho_{H_2}}$ = 10$^{2.5\pm0.1}$ cm$^{-3}$, and a high-excitation component with T$_\mathrm{kin}$ = 40$\pm$5 K and $\mathrm{\rho_{H_2}}$ = 10$^{4.1\pm 0.3}$ cm$^{-3}$ (comparable to the dust temperature of 40 K found by \citealt{kovacs2006})\footnote{We explored the case where the CO(6-5) line luminosity is contaminated by continuum emission by up to 50\%. In the most extreme case, both the Akaike and Bayesian information criteria favour a one-component model.}. Despite the significantly revised CO(1-0) luminosity and linewidth, these parameters are almost identical to those found by \citetalias{riechers2011}, with the only difference being the density of the warm component, where we find 10$^{4.1}$ cm$^{-3}$ compared to 10$^{4.3}$ cm$^{-3}$ in \citetalias{riechers2011}.

\medskip
\section{Discussion}\label{sec:discussion}

\medskip
\subsection{Comparison with the literature}\label{reservoir_context}
CO lines from levels higher than J=2 are poor tracers of the total molecular gas content in galaxies due to their excitation requirements. Some studies of SMGs (including SMM J13120) have found evidence for very extended CO(1-0) reservoirs and/or broad linewidths in some cases (\citealt{ivison2011}, \citetalias{riechers2011}) which are not observed in the higher-J CO transitions. Meanwhile, observations sensitive to low-surface-brightness emission on even larger angular scales have uncovered such massive gas reservoirs with radii of tens of kiloparsecs in a range of systems at high-z, from the Spiderweb Galaxy \citep{emonts2016} to dusty star forming galaxies and hyper-LIRGs \citep{frayer2018,emonts2019}. These studies suggest the existence of a diffuse, low excitation gas component that is not apparent from mid/high-J observations. 

Despite its narrow linewidth, the total cold gas reservoir inferred from the CO(1-0) emission in SMM J13120 has a mass of (13 $\pm$ 3) $\times$ 10$^{10}$ M$_\odot$, and is extended over $\sim$16 kpc in diameter. 
Although the revised CO(1-0) luminosity is significantly lower than previous estimates \citepalias{riechers2011}, we additionally find that the CO SLED still suggests the presence of two gas components to account for all the recovered CO(1-0) emission, including a diffuse component with T$_\mathrm{kin}$ = 25$\pm$5 K and a warmer, denser component with T$_\mathrm{kin}$ = 40$\pm$5 K.
These findings are in line with other studies of high-z systems with available low-J CO detections (see e.g., \citealt[][for unlensed galaxies]{ivison2011,riechers2011, hodge2013}, and \citealt[][for lensed sources]{yang2017,Canameras2018,harrington2020}), reinforcing the importance of low-J CO observations for establishing the nature of the molecular gas reservoirs in high-$z$ systems and their role in early galaxy evolution.

Due to the difficulty of detecting and spatially resolving CO(1-0) at high redshift, there are few other resolved observations of low-J CO transitions in SMGs in the literature. The most direct comparison can be made with GN20 \citep{carilli2010gn20sled,hodge2012,hodge2015}, an unlensed, z = 4.05 SMG, and the only other such source with kpc-scale low-J CO imaging. In particular, extensive high-resolution (0.2$''$/1.3$\sim$kpc) observations with the JVLA in CO(2-1) revealed a clumpy, almost equally-extended cold molecular gas reservoir \citep[$\sim$14 kpc in diameter;][]{hodge2012}. However, contrary to the study presented here, a dynamical analysis of GN20 found evidence of ordered disk rotation. We note that wet mergers have been shown to rapidly relax into smoothly rotating disks and/or be observationally indistinguishable from rotationally supported disks \citep[e.g.,][]{robertson2006,robertson_bullock2008,bornaud2009}
and thus the presence of ordered rotation alone does not automatically preclude a major merger origin. Nevertheless,
the present high-resolution CO(1-0) observations of SMM J13120 stand in stark contrast to the disk-like rotation revealed by the high-resolution CO(2-1) imaging of GN20, implying that we may have caught SMM J13120 in a distinct and potentially short-lived evolutionary stage, as discussed further in the following Section.

\medskip
\subsection{The fate of SMM J13120} \label{sec5.2}
With the wealth of data available for SMM J13120, we are now able to speculate about the evolutionary state of this massive, gas-rich galaxy. Previous authors suggested that a late-stage merger could be the most likely explanation for the morphology and velocity field structure in SMM J13120 \citep{hainline2006,engel2010,riechers2011}. In the light of our data, we reexamine the evidence supporting this scenario.

SMM J13120 shows a complex and discrepant morphology in CO(1-0) and CO(6-5) emission. The CO(1-0) 0th moment map (Figure \ref{moment_maps}, top) shows two distinct components separated by $\sim$11 kpc. The elongated shape of the main component could be indicative of molecular gas being redistributed around a central potential well. The CO(6-5) emission shows flattening almost perpendicular to CO(1-0), indicating a significant difference in the distribution of diffuse and dense molecular gas.

The narrow linewidth seems at odds with other mergers, which report FWHM $\sim$1000 km s$^{-1}$ \citep[e.g.,][]{oteo2016}, and might be a priori taken as a signature of a dynamically cold, ordered system. However, the PV diagram (Figure \ref{pv-diagram}) and the resolved velocity maps (Figure \ref{moment_maps}), as well as the low dynamical mass estimate, indicate that the source is viewed close to face-on. In particular, the 1st moment maps show that the velocity gradient follows the minor axis of CO(1-0) emission. The velocity gradient can also be observed in the PV diagram of the CO(6-5) emission. This kinematic structure in the CO(1-0) emission rules out a circular, regularly rotating disk model for the cold molecular gas reservoir, and lends further support to the merger scenario.

These velocity gradients could indicate, however, that the molecular gas, which is at this point mostly concentrated in the main component,  is starting to settle into a rotating disc. Simulations have shown that rotating discs of molecular gas can reform on very short timescales after a merger \citep{robertson2004,volker2005,robertson2006,hopkins2009}. If SMM J13120 is indeed a late-stage merger, the process of resettling could have already started. The fact that the velocity gradient is aligned with the minor axis is harder to explain. However, CO(6-5) emission is rotated $\sim$70$\degree$ with respect to CO(1-0), such that the observed velocity gradient (pictured at 47$\degree$) more closely corresponds to the major axis of the CO(6-5) emission. This suggests that the disturbed global CO(1-0) morphology may be due to gas that is still being redistributed, which affected the derivation of the major and minor axis beyond what would be expected from a classical disk.

\begin{figure}
    \centering
    \includegraphics[width=\linewidth]{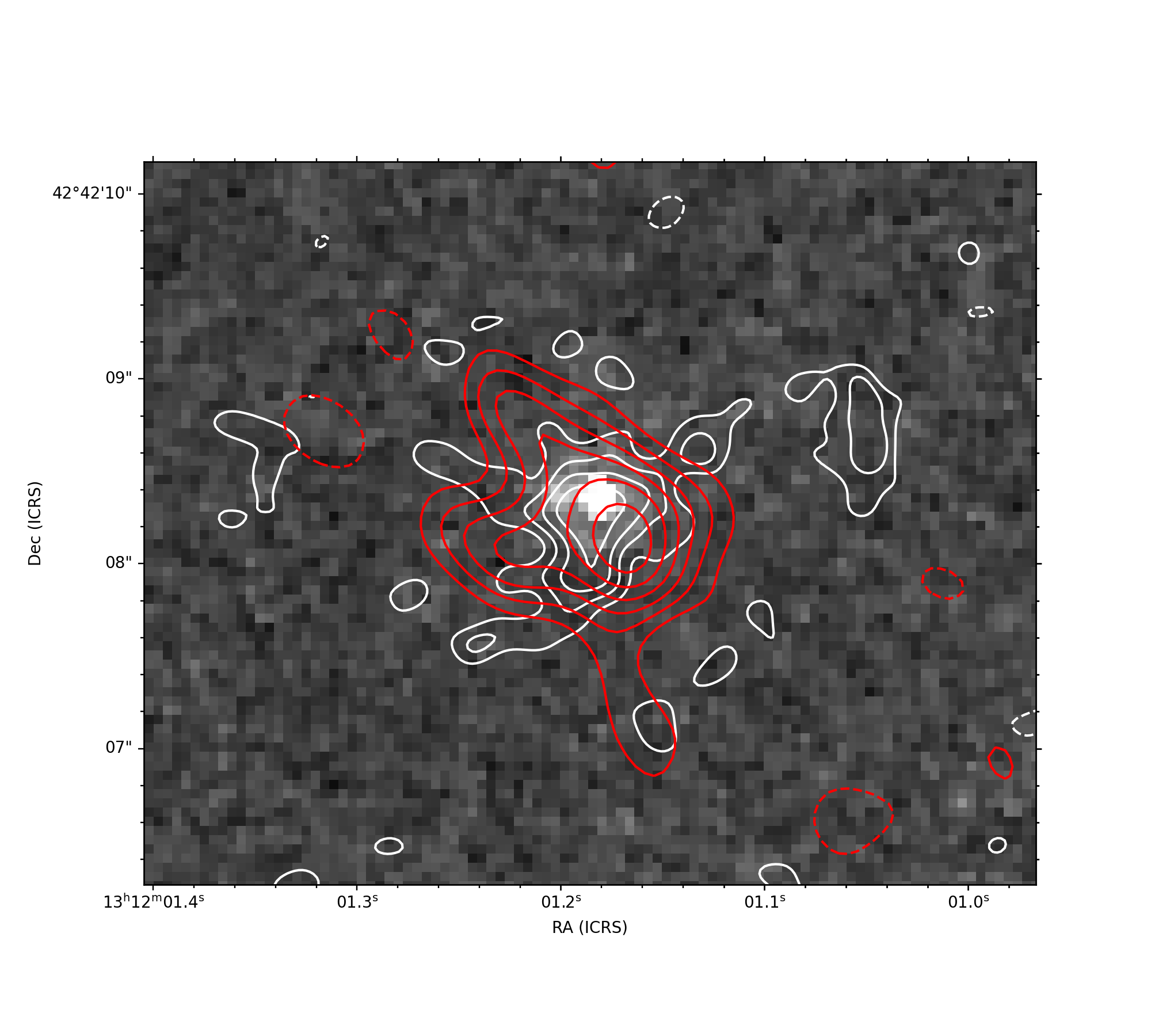}
    \caption{HST-NICMOS F160W-band image with CO(1-0) (white) and CO(6-5) (red) contours overplotted on top. Contours start at $\pm$ 2$\sigma$ and increase in steps of $\sigma$ = 4.4 and 93.8 $\mu$Jy beam $^{-1}$ km s$^{-1}$. The beam sizes are 0.39$"$ and 0.6$"$ for CO(1-0) and CO(6-5), respectively. The stellar emission is concentrated and its peak is coincident with the CO(1-0) and CO(6-5) emission peaks. CO(6-5) emission is equally extended as CO(1-0), and its semi major axis appears to be almost perpendicular to that of CO(1-0).}
    \label{hst}
\end{figure}

Finally, the merger scenario is supported by a comparison between the CO(1-0) and CO(6-5) line emission and near-IR imaging. Figure \ref{hst} shows the near-IR HST/ACS F160W (rest-frame $\lambda\sim$ 0.4$\mu$m) image of SMM J13120. We updated the astrometry of the F160W image by fitting the sources detected in the same field with elliptical Gaussians and matching the peak flux positions with those listed in the Hubble Source Catalogue \footnote{https://catalogs.mast.stsci.edu/hsc/}. 
The CO(6-5) emission is extended on similar scales as CO(1-0), and the peak emission of both lines is coincident with the stellar emission in SMM J13120. The NW component seen in CO(1-0) is not detected in CO(6-5) and near-IR. The compact stellar distribution \citep[r = 1.3$\pm$0.4 kpc,][]{swinbank2010} suggests that the merger is in a late stage. 

In addition to its large submillimiter flux and extended cold gas reservoir, SMM J13120 shows emission lines in the optical spectrum indicative of the presence of an AGN \citep{chapman2005}. This system appears thus to present characteristics of both the SMG and AGN populations. According to the current paradigm of galaxy evolution \citep{sanders1988,hopkins2008b,stacey2018}, a fraction of the AGN population at high-z is transitioning from merger-driven SMGs to unobscured quasars. This transitioning population would retain high rates of dust-obscured star formation, thus being luminous in the far-infrared to mm regime. SMM J13120 presents all the characteristics expected for this population. It is thus possible that SMM J13120 is currently transitioning between an SMG and unobscured QSO. The low depletion timescale suggests a phase of rapid star formation that will quickly deplete its gas reservoir, leaving the SMG phase and becoming an AGN.
Our SED fitting reveals that, with SFR = 948$^{+420}_{-418}$ M$_\odot$ yr$^{-1}$ and M$_*$ = 6.5$^{+2.5}_{-1.9}\times$10$^{11}$ M$_\odot$, SMM J13120 sits on the massive end of the main sequence at z=3.4 \citep[]{speagle2014}. However, the depletion time of 137 Myr is a factor of ~2-3 lower than those observed in the general SMG population at this redshift (200-300 Myr, \citealt{bothwell2013,tacconi2018,Dudzeviciute2020,birkin2020}). With a gas fraction also four times lower than expected from empirical scaling relations \citep{tacconi2018}, this main-sequence galaxy presents characteristics typical of starbursts. \citet[]{elbaz2018} found four z$\sim$2 Herschel-selected MS galaxies to be starbursts `hidden' in the main sequence, with short depletion times, low gas fractions and a spheroid-like morphology in the H-band, attributes we also find in SMM J13120. 

The case of SMM J13120 therefore raises two important, connected points. First, it reinforces previous findings that mergers may drive important physical transformations of galaxies without pushing them off the main sequence, although most of the currently available work has been done on the most massive galaxies of the main sequence \citep[M$_*$ $>$ 10$^{11}$ M$_\odot$;][]{fensch2017,silva2018,cibinel2019MNRAS,puglisi2019,drew2020}. Second, it demonstrates that the massive end of the main sequence hosts a variety of star formation modes, rather than solely `typical star-forming galaxies' driven by secular evolution. 

\medskip
\section{Conclusion} \label{sec:conclusion}

We have presented high-resolution ($0.39"$) multi-configuration VLA observations of the z=3.4 SMG SMM J13120+4242 in the CO(1-0) emission line. We derive a gas mass of M$_{\text{gas}}$= (13 $\pm$ 3) $\times$ 10$^{10}$ ($\alpha_\mathrm{CO}$/1.0) M$_\odot$. The molecular gas reservoir, imaged on $\sim$3 kpc scales, extends over $>$16 kpc, in agreement with previous measurements. The spectral line profile reveals a FWHM of 267 $\pm$ 64 km s$^{-1}$. The deeper data presented in this study are quantitatively consistent with previous, less sensitive VLA CO(1-0) observations, and rule out the presence of a broader linewidth component which appeared previously at low SNR.

We use a virial estimator to estimate a dynamical mass of M$_{\text{dyn}}$ = (1.6 $\pm$ 0.9) $\times$ 10$^{11}$ M$_\odot$. This value is lower than the stellar mass derived from SED fitting (6.45 $\times$ 10$^{11}$ M$_\odot$), highlighting the impact of the non-gravitationally supported motions seen here on the derivation of the dynamical mass estimate. We thus treat it as a lower limit and use it in combination with the stellar mass to put an upper limit on the CO-to-H$_2$ mass conversion factor, $\alpha_\mathrm{CO}$ $<$ 1.4 M$_\odot$ K$^{-1}$ km$^{-1}$ s pc$^{-2}$, consistent with a ULIRG-like value and comparable with the values derived for other high-z SMGs.

The massive, extended molecular gas reservoir, with $>$16 kpc diameter, shows a complex velocity field with evidence for a velocity gradient. However, the velocity gradient aligns approximately with the minor axis of the CO(1-0) reservoir, precluding the modelling of the velocity field as a circular rotating disk. Using previous observations of CO(6-5), we confirm that the observed velocity gradient is seen in both transitions, but appears to align with the major axis of the CO(6-5) emission. This suggests that the perturbed gas may have already started settling into a disk but has not yet circularised. A late-stage intermediate (5:1) merger is the most likely cause for the disturbed morphology, chaotic velocity field and intense burst of star formation taking place in SMM J13120. 

The CO excitation ladder of the system is well fit by a two component ISM model, with a cold, diffuse component and a warm, denser component, as is characteristic of the average SMG population. The properties presented in this paper suggest that SMM J13120 could be in a transitioning phase between a merger-driven starburst and an unobscured QSO. Given its low depletion timescale ($\sim$100 Myr), SMM J13120 is expected to quickly deplete its gas reservoir, eventually turning into an unobscured, gas-poor QSO.

We highlight the importance of high-resolution, low-J CO studies for characterising the ISM in high-z SMGs. Although resolved imaging of CO(1-0) at high-z requires significant time investments, our case study of SMM J13120 demonstrates that CO(1-0) is key to accurately characterise the dynamical state of the cold molecular gas. High-resolution kinematic studies like this one are also crucial to better understand differences between gas phases of different excitation. Future surveys of the CO(1-0) gas in larger samples of SMGs and their high-resolution follow-up will shed light on the typical molecular gas conditions in these sources. 

\acknowledgements{
We thank the referee for valuable comments that have improved the analysis presented in this manuscript. We thank H. Engel and R. Davies for kindly providing PdBI CO(6-5) observations of SMM J13120. M.F.C., J.A.H., and M.R. acknowledge support of the VIDI research programme with project number 639.042.611, which is (partly) financed by the Netherlands Organisation for Scientific Research (NWO). M.R. is supported by the NWO Veni project “Under the lens” (VI.Veni.202.225). D.R. acknowledges support from the National Science Foundation under grant numbers AST-1614213 and AST-1910107. D.R. also acknowledges support from the Alexander von Humboldt Foundation through a Humboldt Research Fellowship for Experienced Researchers. The National Radio Astronomy Observatory is a facility of the National Science Foundation operated under cooperative agreement by Associated Universities, Inc.}

\bibliography{sample63}{}
\bibliographystyle{aasjournal}

\appendix\label{sec:appendix}

\begin{figure}[h]
    \centering
    \includegraphics[scale=0.7]{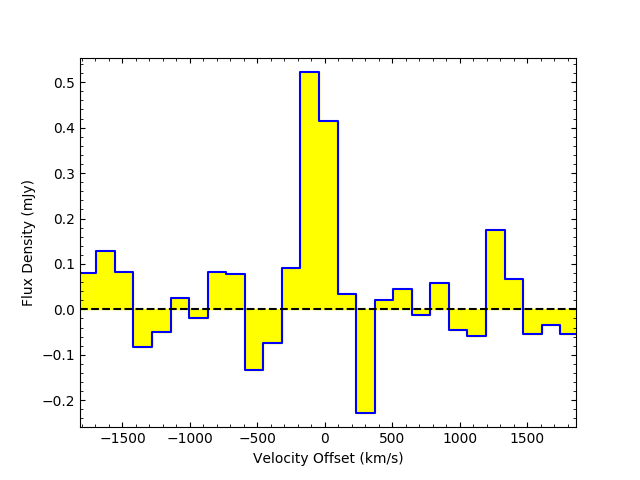}
    \caption{CO(1-0) spectrum of SMM J13120 with a spectral resolution of 138 km s$^{-1}$. No evidence for a broad ($\sim$1000 km s$^{-1}$) line is seen in the deeper multi-configuration data.}
    \label{linewidth6}
\end{figure}

With the earlier VLA CO(1-0) data set used in \citetalias{riechers2011}, the CO(1-0) spectrum indicated a very broad linewidth of $\sim$ 1000 km s$^{-1}$, similar to the GBT single-dish spectrum from \citealt{hainline2006}. After independently re-analysing the \citetalias{riechers2011} data, we recover the seemingly extended line emission. However, the deeper multi-configuration data presented here (including the \citetalias{riechers2011} data) reveals a much narrower line FWHM. Figure \ref{linewidth6} shows the CO(1-0) spectrum of SMM J13120 binned in channels of 138 km s$^{-1}$ to increase the SNR. We do not recover any such broad component despite the fact that the combined data are two times more sensitive at matched resolution. 

\begin{figure*}
    \centering
    \includegraphics[scale=0.4]{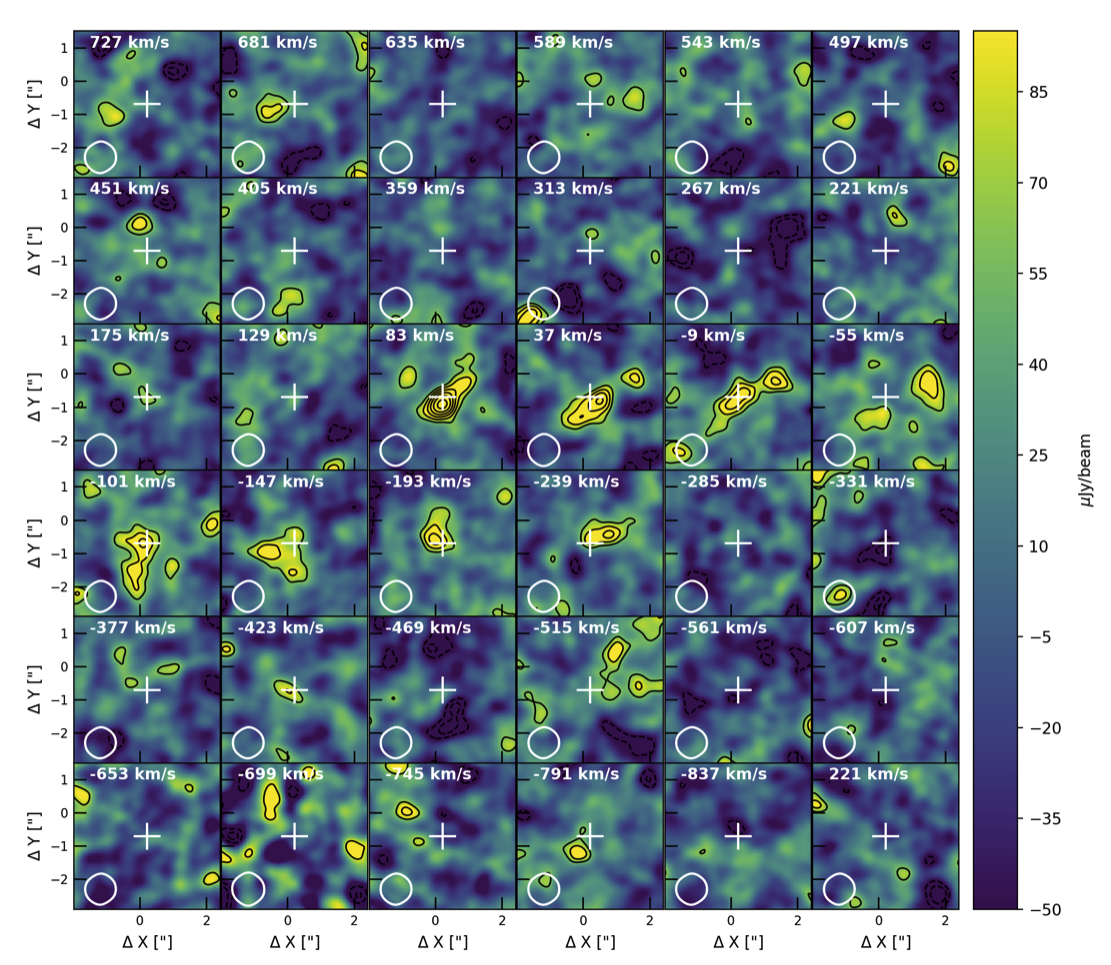}
    \caption{CO(1-0) emission maps in SMM J13120 in 46 km s$^{-1}$ channels. The data was tapered to a resolution of 0.94$"$ to search for faint extended emission that could have been missed at higher resolution. The shape of the beam is displayed in each channel. The white cross corresponds to the best fit peak position of a 2D elliptical Gaussian to the 0th moment map. Contours start at $\pm$2$\sigma$ and increase in steps of $\sigma$ = 35 $\mu$Jy beam$^{-1}$.}
    \label{tapered_chans}
\end{figure*}

\begin{figure*}
\centering
    \includegraphics[scale=0.4]{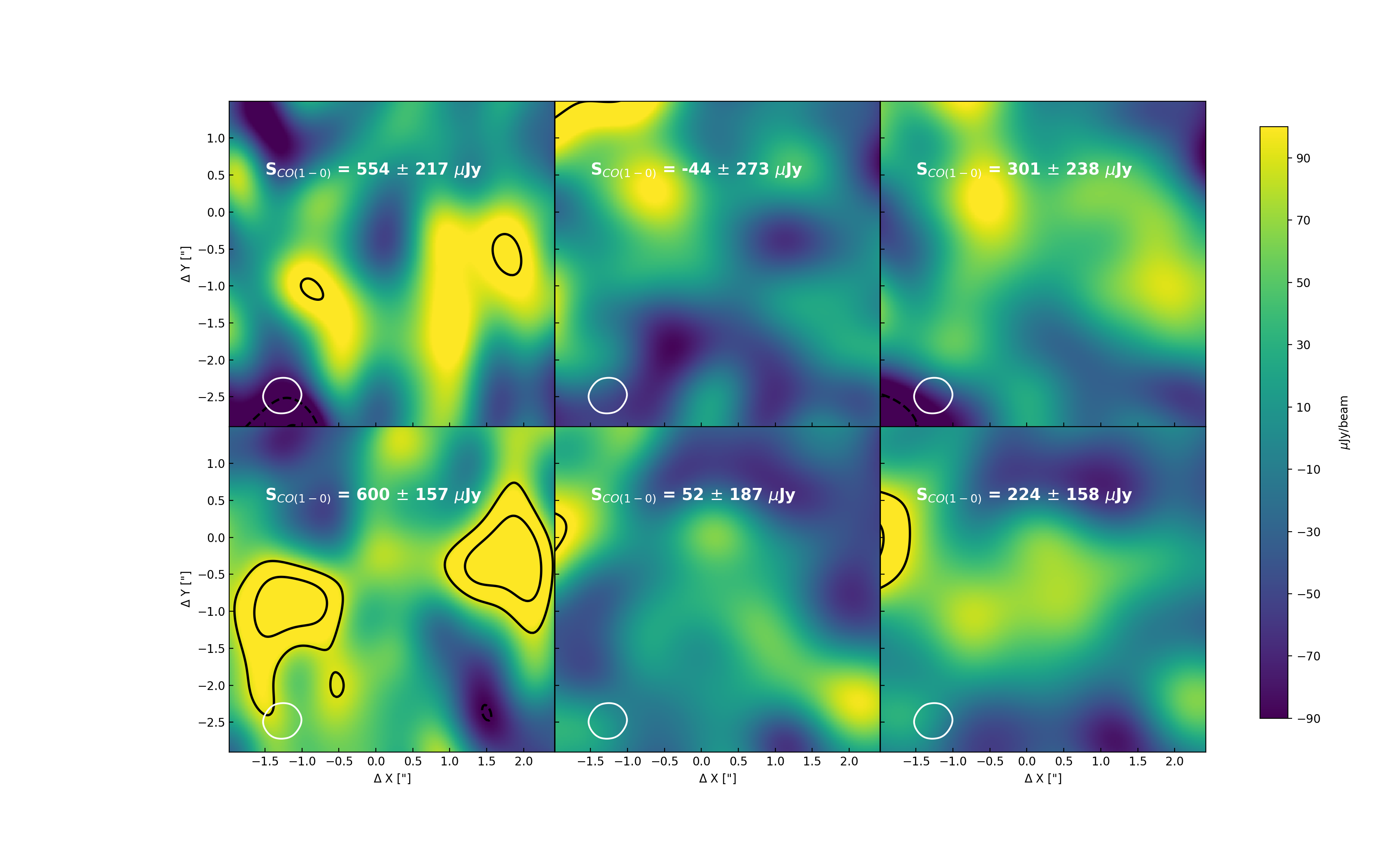}
    \caption{0th moment maps collapsed over the negative and positive velocity offsets where emission was present in the spectrum in \citetalias{riechers2011}. The top and bottom rows show the emission at negative (-800 to -600 km s$^{-1}$) and positive (400 to 600 km s$^{-1}$) velocity offsets, respectively. The first, second and third column correspond to the earlier VLA dataset \citepalias{riechers2011}, the new dataset presented here and the concatenated data, respectively. The flux density was extracted from the same circular aperture. The new and concatenated datasets do not show emission above the 2$\sigma$ noise, and the flux density is consistent with the earlier VLA data within the errors, highlighting the uncertainty in the original dataset.}
    \label{wings}
\end{figure*}

To further search for this broad component in the new data, we tapered both sets to the same resolution of 0.94$"$.  The tapered channel maps do not show further emission with respect to the native resolution channel maps (Figure \ref{tapered_chans}). We then made 0th moment maps collapsed over the negative (-800 to -600 km s$^{-1}$) and positive (400 to 600 km s$^{-1}$) velocity offsets where emission was originally apparent in the spectrum in \citetalias{riechers2011}. The 0th moment maps are shown in Fig. \ref{wings}. We compare emission seen in the earlier VLA data \citetalias{riechers2011}, the new data set and the combined datasets. We do not recover any emission above 2$\sigma$ significance in the new and combined datasets. Moreover, the flux densities we measure are consistent within 2$\sigma$ with the earlier \citetalias{riechers2011} data, highlighting the uncertainty in the original dataset.

\end{document}